\documentclass[10pt,a4paper,fleqn]{article}
\usepackage[margin=25mm]{geometry}
\usepackage[T1]{fontenc}
\usepackage{lmodern}
\usepackage{microtype}
\usepackage[numbers,sort&compress]{natbib}
\usepackage{graphicx}
\usepackage{amsmath,amssymb,amsfonts,mathrsfs,bm}
\usepackage{xcolor,float,placeins,soul,booktabs,caption,etoolbox}
\usepackage{hyperref}
\providecommand{\upi}{\mathrm{\pi}}
\hypersetup{colorlinks=true,allcolors=blue,pdfauthor={Peishuo Wu and Chi Zhu},pdftitle={A variational multiscale rotational pressure-correction method for incompressible flows with open boundaries}}
\graphicspath{{figs/}}
\AtBeginEnvironment{tabular}{\rmfamily\small}
\title{A variational multiscale rotational pressure-correction method for incompressible flows with open boundaries}
\author{Peishuo Wu \and Chi Zhu\thanks{Corresponding author. Email: \href{mailto:chi.zhu@pku.edu.cn}{chi.zhu@pku.edu.cn}.}}
\date{}
\begin{document}
\maketitle
\begin{center}
\small School of Mechanics and Engineering Science,\\
State Key Laboratory for Turbulence and Complex Systems,\\
Peking University, Beijing 100871, China
\end{center}
\begin{abstract}
Rotational pressure correction improves pressure accuracy near open boundaries in incompressible flow simulations, but equal-order finite element discretizations can still develop spurious pressure oscillations. We develop a residual-based variational multiscale rotational pressure-correction scheme (RPCS-VMS) that stabilizes both the pressure-increment and pressure-recovery equations. By constructing stage-specific momentum residuals using a common intermediate-velocity acceleration, the scheme retains four sequential linear solves per time step. For a reduced BDF1 Oseen model with frozen stabilization coefficients, we establish augmented-energy decay and a conditional bound on the pressure gradient. Manufactured-solution tests demonstrate suppression of pressure oscillations at small time steps and reduced near-boundary pressure errors. With \(P_1/P_1\) and \(Q_1/Q_1\) elements, the observed spatial \(L^2\) convergence rates are 2 for velocity and approximately \(1.5\) for pressure, while first- and second-order backward differentiation formulas attain their expected temporal orders. Branched-pipe and two- and three-dimensional cylinder benchmarks assess outlet flow partitioning, transient forces, and pressure differences, with results agreeing closely with reference solutions. In the three-dimensional cylinder test, RPCS-VMS achieves a 2.41-fold wall-time speedup over the tested monolithic VMS solver.
\end{abstract}


\section{Introduction}

Pressure-correction methods advance the incompressible Navier-Stokes equations through separate velocity and pressure solves, avoiding a coupled saddle-point system at each time step. Following Chorin's projection method \citep{Chorin1968Numerical}, fractional-step formulations have been developed for a range of spatial discretizations and time integration orders \citep{Goda1979Multistep,KimMoin1985Fractional,vanKan1986Second,Karniadakis1991High,Guermond2006Overview}. Their pressure accuracy depends in part on the compatibility of the auxiliary boundary conditions with the coupled problem \citep{StrikwerdaLee1999Accuracy,Brown2001Accurate,Guermond2006Overview}. When equal-order finite elements are used, control of discrete pressure modes is a further requirement. Boundary accuracy and pressure stability therefore need separate consideration.

The rotational pressure-correction scheme (RPCS), introduced by Timmermans, Minev, and Van de Vosse \citep{Timmermans1996Approximate}, modifies the standard incremental pressure update by incorporating a correction based on the divergence of the intermediate velocity. Guermond and Shen \citep{Guermond2004On} subsequently established improved velocity and pressure error estimates for this scheme under the assumptions of their analysis. At open boundaries, prescribed data couple velocity derivatives and pressure, making their treatment in the auxiliary subproblems important to splitting accuracy. For time-dependent Stokes problems, Guermond, Minev, and Shen \citep{Guermond2006Error} identified limitations of standard pressure correction and established error estimates for the rotational formulation. Subsequent work developed several approaches to imposing open and traction boundary data. Liu \citep{Liu2009Open} proposed pressure boundary conditions within a pressure-Poisson formulation, while Poux et al.~\citep{Poux2011Improvements,Poux2012Open} investigated improved boundary enforcement for pressure- and velocity-correction schemes. B\"ansch \citep{Bansch2014Finite} modified the traction condition for the intermediate velocity through a boundary correction involving the pressure increment. Bonito, Guermond, and Lee \citep{Bonito2015Modified} instead modified the pressure-increment boundary condition to reduce the pressure boundary layer in a first-order Stokes scheme.

Backflow introduces an additional difficulty because the convective boundary flux can supply kinetic energy to the computational domain. Energy estimates and open-boundary treatments addressing this issue have been developed by Bruneau and Fabrie \citep{BruneauFabrie1994Effective,BruneauFabrie1996New} and Dong and Shen \citep{Dong2015Pressure}. Becker et al.~\citep{Becker2015Finite} developed a general weak boundary formulation with discrete kinetic-energy control. For rotational pressure correction with open or traction boundaries, Lee and Salgado \citep{LeeSalgado2016Stability} established stability results for grad-div-stabilized and boundary-correction schemes. These developments provide important guidance on boundary accuracy and energy behavior. For equal-order discretizations, the resulting split formulation must also provide control of discrete pressure modes.

Equal-order velocity-pressure interpolation simplifies implementation, but standard continuous low-order pairs generally lack the inf-sup stability of pairs such as Taylor-Hood elements \citep{Brezzi1974On,Taylor1973Numerical}. Established remedies include residual-based stabilization \citep{Hughes1986New,Franca1992Stabilized,Tezduyar1992Incompressible}, pressure-gradient and projection methods \citep{Codina1997Finite,Becker2001Finite,Bochev2006Stabilization}, and interior-penalty terms \citep{Burman2006Edge,Burman2007Continuous}. Some fractional-step schemes also supply pressure control, but it depends on the time step and can weaken as the step decreases \citep{Codina2001Pressure,BadiaCodina2007Convergence}. This motivates additional spatial stabilization for pressure approximation at small time steps. Such stabilization can itself perturb the pressure near physical boundaries \citep{Becker2001Finite}, and these perturbations must be distinguished from boundary errors introduced by splitting.

The variational multiscale (VMS) framework relates stabilization to the modeled influence of unresolved scales on the finite element solution \citep{Hughes1995Multiscale,Hughes1998Variational}, with different subscale models determining the resulting stabilization terms \citep{Codina2000Stabilization,Codina2002Stabilized,Codina2007Time,Masud2006Multiscale,Codina2017Variational}. Its use with pressure segregation is established. Codina and Soto \citep{Codina2004Approximation} combine orthogonal-subscale stabilization with an iterative pressure-segregation procedure whose converged solution corresponds to the monolithic time discretization. Codina and Badia \citep{Codina2006On} examine fractional-step and other segregation algorithms and establish stability estimates for fractional-step schemes without an inf-sup assumption. They also identify the weakness of the resulting pressure estimates and explain that stronger control can be obtained from inf-sup stability or spatial stabilization.

In rotational pressure correction, the pressure increment and updated pressure are determined by different equations linked through the rotational update. This raises the question of how to coordinate residual stabilization across these stages while retaining sequential solves. The resulting formulation also requires an assessment of pressure control, since energy decay alone need not exclude discrete pressure oscillations.

We develop a residual-based VMS rotational pressure-correction scheme (RPCS-VMS) that coordinates stabilization in the pressure-increment and pressure-recovery equations while retaining four sequential solves. For a reduced homogeneous BDF1 Oseen model on a fixed $P_1/P_1$ mesh with frozen coefficients, we establish augmented-energy decay under stated coefficient conditions and a conditional bound on the pressure gradient. Manufactured-solution tests and flow benchmarks assess convergence, pressure behavior, and computational cost.

This paper is organized as follows. Sections~\ref{sec:rpcs} and~\ref{sec:increment-bc} present the baseline split and examine its boundary compatibility. Section~\ref{sec:vms} develops RPCS-VMS, and Section~\ref{sec:stability} states the reduced-model energy and pressure-gradient estimates. Sections~\ref{sec:results} and~\ref{sec:conclusions} present the numerical results and conclusions, respectively.

\section{Baseline Rotational Pressure-Correction Framework}
\label{sec:rpcs}

Here, we summarize the governing equations and the classical rotational pressure-correction scheme that provide the baseline for the boundary treatment in Section~\ref{sec:increment-bc} and the VMS formulation in Section~\ref{sec:vms}.

\subsection{Governing Equations and Boundary Conditions}

Let \(\Omega\subset\mathbb R^d\), \(d=2\) or \(3\), be a bounded, connected Lipschitz domain. The boundary \(\Gamma=\partial\Omega\) comprises a prescribed-velocity portion \(\Gamma_D\) and an open-boundary portion \(\Gamma_N\), with \(\Gamma_D\cap\Gamma_N=\emptyset\) and \(\overline{\Gamma_D}\cup\overline{\Gamma_N}=\Gamma\). The outward unit normal is denoted by \(\bm n\). The incompressible Navier-Stokes problem is
\begin{align}
    \partial_t\bm u+(\bm u\cdot\nabla)\bm u-\nu\nabla^2\bm u+\nabla p
    &=\bm f,
    &&\text{in }\Omega\times(0,T],
    \label{eq:strong-momentum}\\
    \nabla\cdot\bm u
    &=0,
    &&\text{in }\Omega\times(0,T],
    \label{eq:strong-continuity}\\
    \bm u
    &=\bm u_D,
    &&\text{on }\Gamma_D\times(0,T],
    \label{eq:strong-dirichlet}\\
    \nu\partial_n\bm u-p\bm n
    -\frac12(\bm u\cdot\bm n)_-\bm u
    &=\bm h,
    &&\text{on }\Gamma_N\times(0,T],
    \label{eq:obc}\\
    \bm u(\cdot,0)
    &=\bm u_0,
    &&\text{in }\Omega.
    \nonumber
\end{align}
Here, \(T>0\) is the final time, and \(\bm u_0\) the initial velocity. The constant \(\nu>0\) is the kinematic viscosity, \(p\) is the pressure divided by the constant fluid density, and \(\bm f\) and \(\bm h\) use the corresponding density-normalized units. We write \((\bm u\cdot\bm n)_-:=\min\{\bm u\cdot\bm n,0\}\). This backflow term is motivated by Bruneau and Fabrie \citep{BruneauFabrie1994Effective,BruneauFabrie1996New} and related energy-stable open-boundary formulations \citep{Dong2014Robust,Dong2015Pressure}. It acts only where \(\bm u\cdot\bm n<0\), controlling the adverse energy contribution of flow entering through an open boundary. The Laplacian viscous operator gives the natural boundary quantity \(\nu\partial_n\bm u-p\bm n\), termed the pseudo-traction. The same Laplacian/pseudo-traction pairing is used in Refs.~\citep{Guermond2006Overview,Dong2015Pressure}.

\subsection{Classical Rotational Pressure-Correction Scheme}

Our development builds on the classical RPCS introduced by Timmermans, Minev, and Van de Vosse \citep{Timmermans1996Approximate}. Let \(t_n=n\Delta t\), with uniform time step \(\Delta t\). For backward differentiation formulas (BDF), define the history by \(\bm U^n=\sum_{j=1}^{J}\alpha_j\bm u^{n+1-j}\), so that the discrete time derivative is \((\gamma_0\bm u^{n+1}-\bm U^n)/\Delta t\). The first- and second-order schemes, denoted BDF1 and BDF2, use the following coefficients, history, and explicit advecting velocity:
\begin{equation}
    \begin{aligned}
        \text{BDF1:}\qquad
        &J=1,\qquad \gamma_0=1,\qquad \alpha_1=1,\\
        &\bm U^n=\bm u^n,\qquad
        \bm u^*=\bm u^n,\\[3pt]
        \text{BDF2:}\qquad
        &J=2,\qquad \gamma_0=\frac32,\qquad
        (\alpha_1,\alpha_2)=\left(2,-\frac12\right),\\
        &\bm U^n=2\bm u^n-\frac12\bm u^{n-1},\qquad
        \bm u^*=2\bm u^n-\bm u^{n-1}.
    \end{aligned}
    \label{eq:bdf-history-notation}
\end{equation}
BDF2 computations are initialized with one BDF1 step. The extrapolated field \(\bm u^*\) is fixed during each time step. Given the required velocity history and \(p^n\), the update consists of four steps.

\paragraph{Step 1: Intermediate velocity.}
Compute \(\tilde{\bm u}^{n+1}\) from
\begin{equation}
    \frac{\gamma_0\tilde{\bm u}^{n+1}-\bm U^n}{\Delta t}
    +(\bm u^*\cdot\nabla)\tilde{\bm u}^{n+1}
    -\nu\nabla^2\tilde{\bm u}^{n+1}
    +\nabla p^n
    =\bm f^{n+1},
    \qquad\text{in }\Omega.
    \label{eq:predictor-strong}
\end{equation}

\paragraph{Step 2: Pressure increment.}
The pressure increment \(\phi^{n+1}\) is obtained from
\begin{equation}
\frac{\Delta t}{\gamma_0}\nabla^2\phi^{n+1}
    =
\nabla\cdot\tilde{\bm{u}}^{n+1},
\qquad \text{in }\Omega.
\label{eq:phi-poisson}
\end{equation}

\paragraph{Step 3: Velocity correction.}
The corrected velocity is obtained through
\begin{equation}
\bm{u}^{n+1}
    =
\tilde{\bm{u}}^{n+1}
-\frac{\Delta t}{\gamma_0}\nabla\phi^{n+1},
\qquad \text{in }\Omega.
\label{eq:velocity-correction}
\end{equation}
Equations~\eqref{eq:phi-poisson} and~\eqref{eq:velocity-correction} imply
\(
\nabla\cdot\bm u^{n+1}=0
\)
at the continuous level.

\paragraph{Step 4: Rotational pressure recovery.}
The pressure is updated as
\begin{equation}
    p^{n+1}
    =
    p^n+\phi^{n+1}
-\nu\nabla\cdot\tilde{\bm{u}}^{n+1},
\qquad \text{in }\Omega.
\label{eq:rotational-pressure-update}
\end{equation}

\section{Boundary Conditions and Compatibility}
\label{sec:increment-bc}

Section~\ref{sec:rpcs} specifies the interior equations of the RPCS. We now prescribe boundary conditions for the intermediate velocity and pressure increment and examine their compatibility with the boundary data for the corrected fields. The discussion is conducted at the space-continuous, time-discrete level.

\subsection{Open Boundary}

On \(\Gamma_N\), we freeze the backflow coefficient \(\beta^*:=(\bm u^*\cdot\bm n)_-\) during each step to keep the boundary term linear. The target boundary relation for the corrected velocity and updated pressure is the following semi-implicit form of Eq.~\eqref{eq:obc}, motivated by energy-stable pressure-correction formulations \citep{Dong2015Pressure}:
\begin{equation}
    \nu\partial_n\bm u^{n+1}
    -p^{n+1}\bm n
    -\frac12\beta^*\bm u^{n+1}
    =\bm h^{n+1},
    \qquad \text{on }\Gamma_N .
    \label{eq:semi-implicit-obc}
\end{equation}
At the intermediate-velocity stage, the updated pressure is not yet available. We therefore use the known pressure \(p^n\) in the boundary condition, keeping this solve independent of the pressure increment:
\begin{equation}
    \nu\partial_n\tilde{\bm u}^{n+1}
    -p^n\bm n
    -\frac12\beta^*\tilde{\bm u}^{n+1}
    =\bm h^{n+1},
    \qquad \text{on }\Gamma_N .
    \label{eq:velocity-prediction-obc}
\end{equation}

We next seek the condition on \(\phi^{n+1}\) needed to recover the normal component of Eq.~\eqref{eq:semi-implicit-obc}. Substitute the velocity correction~\eqref{eq:velocity-correction} and rotational pressure update~\eqref{eq:rotational-pressure-update} into this relation and take its normal component. Subtracting the normal component of Eq.~\eqref{eq:velocity-prediction-obc} then gives
\begin{equation}
    \phi^{n+1}
    +\nu\frac{\Delta t}{\gamma_0}\partial_{nn}\phi^{n+1}
    -\nu\nabla\cdot\tilde{\bm u}^{n+1}
    -\frac{\Delta t}{2\gamma_0}\beta^*\partial_n\phi^{n+1}
    =0,
    \qquad \text{on }\Gamma_N .
    \label{eq:normal-increment-residual}
\end{equation}
Here \(\partial_{nn}q=\bm n^\top(\nabla^2q)\bm n\) for a smooth scalar field \(q\). To obtain a relation involving only \(\phi^{n+1}\), we use Eq.~\eqref{eq:phi-poisson} to write \(\nabla\cdot\tilde{\bm u}^{n+1}=(\Delta t/\gamma_0)\nabla^2\phi^{n+1}\). We assume sufficient regularity to restrict this equation and the second derivatives of \(\phi^{n+1}\) to the boundary. Splitting the Laplacian into normal and surface contributions then allows the second normal derivative in Eq.~\eqref{eq:normal-increment-residual} to cancel.

On a smooth boundary, define the surface gradient \(\nabla_\Gamma q=(\bm I-\bm n\otimes\bm n)\nabla q\), the Laplace-Beltrami operator \(\Delta_\Gamma q=\nabla_\Gamma\cdot\nabla_\Gamma q\), and the curvature \(\kappa=\nabla_\Gamma\cdot\bm n\), where \(\bm I\) is the identity tensor. The Laplacian decomposes as
\begin{equation}
    \nabla^2 q
    =
    \partial_{nn}q
    +\kappa\partial_n q
    +\Delta_\Gamma q .
    \label{eq:surface-laplacian-decomposition}
\end{equation}
Substitution into Eq.~\eqref{eq:normal-increment-residual} yields the surface relation
\begin{equation}
    \phi^{n+1}
    -
    \frac{\Delta t}{\gamma_0}
    \left[
        \nu\Delta_\Gamma\phi^{n+1}
        +
        \left(
            \nu\kappa+\frac12\beta^*
        \right)
        \partial_n\phi^{n+1}
    \right]
    =0,
    \qquad \text{on }\Gamma_N .
    \label{eq:surface-increment-relation}
\end{equation}
This relation expresses normal pseudo-traction compatibility, including tangential diffusion, curvature, and backflow contributions. It does not enforce tangential pseudo-traction matching.

The outlets considered here are planar, so \(\kappa=0\). If backflow is also absent, \(\beta^*=0\), Eq.~\eqref{eq:surface-increment-relation} reduces to
\begin{equation}
    \phi^{n+1}
    -
    \nu\frac{\Delta t}{\gamma_0}
    \Delta_\Gamma\phi^{n+1}
    =0,
    \qquad \text{on }\Gamma_N .
    \label{eq:reduced-surface-increment}
\end{equation}
If the boundary term arising from surface integration by parts vanishes, for example under periodic boundary conditions or homogeneous Neumann conditions on \(\partial\Gamma_N\), testing Eq.~\eqref{eq:reduced-surface-increment} with \(\phi^{n+1}\) gives
\begin{equation}
    \|\phi^{n+1}\|_{L^2(\Gamma_N)}^2
    +
    \nu\frac{\Delta t}{\gamma_0}
    \|\nabla_\Gamma\phi^{n+1}\|_{L^2(\Gamma_N)}^2
    =0.
\end{equation}
Since both terms are nonnegative,
\begin{equation}
    \phi^{n+1}=0,
    \qquad \text{on }\Gamma_N .
    \label{eq:phi-dirichlet-open}
\end{equation}

This conditional argument recovers the conventional homogeneous pressure-increment condition at an open boundary \citep{Guermond2006Overview}, which we adopt for implementation and analysis throughout this study. The above derivation nevertheless clarifies the assumptions under which this condition is exact and identifies the contributions neglected in more general settings.

\subsection{Boundary Conditions for the RPCS Subproblems}

On \(\Gamma_D\), we impose \(\tilde{\bm u}^{n+1}=\bm u_D^{n+1}\). Requiring the velocity correction~\eqref{eq:velocity-correction} to preserve the prescribed normal component then gives \(\partial_n\phi^{n+1}=0\). Together with the adopted open-boundary treatment, this gives the boundary conditions used in the intermediate-velocity and pressure-increment subproblems:
\begin{align}
    \tilde{\bm u}^{n+1}
    &=\bm u_D^{n+1},
    &&\text{on }\Gamma_D,
    \label{eq:velocity-prediction-dirichlet}\\*
    \nu\partial_n\tilde{\bm u}^{n+1}
    -p^n\bm n
    -\frac12\beta^*\tilde{\bm u}^{n+1}
    &=\bm h^{n+1},
    &&\text{on }\Gamma_N,
    \label{eq:rpcs-bc-u-open}\\*
    \partial_n\phi^{n+1}&=0,
    &&\text{on }\Gamma_D,
    \label{eq:phi-neumann-dirichlet}\\*
    \phi^{n+1}&=0,
    &&\text{on }\Gamma_N.
    \label{eq:rpcs-bc-phi-open}
\end{align}
The homogeneous Neumann condition for the pressure increment on \(\Gamma_D\) only preserves the normal component of the prescribed velocity. The full Dirichlet velocity data are enforced in the discrete velocity-correction step in Section~\ref{subsec:vms-implementation}.

\section{RPCS-VMS Formulation}
\label{sec:vms}

Building on the RPCS algorithm and boundary treatment developed in the previous sections, we now construct a residual-based VMS formulation for equal-order velocity-pressure interpolation. The formulation coordinates stabilization of the intermediate-velocity, pressure-increment, and rotational pressure-recovery equations while preserving the four sequential solves. Throughout this section, we omit the superscript \(n+1\) on current unknowns and write \(\delta=\Delta t/\gamma_0\).

\subsection{Momentum Balances for Different Stages}
\label{subsec:vms-scale-decomposition}

We construct three momentum residuals for the VMS stabilization terms in the intermediate-velocity, pressure-increment, and rotational pressure-recovery equations. To make stabilization implicit in the current unknown of each step, we base these residuals on momentum balances involving \(p^n\), \(p^n+\phi\), and \(p\), respectively.

For smooth space-continuous, time-discrete RPCS fields, define the BDF accelerations associated with the intermediate and corrected velocities, together with the provisional pressure, by
\[
 \widetilde{\bm a}=\frac{\gamma_0\tilde{\bm u}-\bm U^n}{\Delta t},
 \qquad \bm a=\frac{\gamma_0\bm u-\bm U^n}{\Delta t},
 \qquad \pi=p^n+\phi.
\]
Both accelerations use the same known velocity history \(\bm U^n\), so their difference comes entirely from the current velocity correction. The first momentum balance is Eq.~\eqref{eq:predictor-strong}, which contains \(\widetilde{\bm a}\) and the known pressure gradient \(\nabla p^n\).

The velocity-correction equation~\eqref{eq:velocity-correction} gives \(\widetilde{\bm a}=\bm a+\nabla\phi\). Substituting this relation into the intermediate-velocity equation yields the second momentum balance associated with the pressure-increment step,
\begin{equation}
 \bm a+(\bm u^*\cdot\nabla)\tilde{\bm u}
       -\nu\nabla^2\tilde{\bm u}+\nabla\pi=\bm f^{n+1}.
\label{eq:vms-continuous-momentum}
\end{equation}

Eq.~\eqref{eq:rotational-pressure-update} gives \(\pi=p+\nu\nabla\cdot\tilde{\bm u}\). Substituting this expression into Eq.~\eqref{eq:vms-continuous-momentum} and using \(\nabla\times\nabla\times\bm v=\nabla(\nabla\cdot\bm v)-\nabla^2\bm v\) gives the third momentum balance associated with the rotational pressure-recovery step,
\begin{equation}
 \bm a+(\bm u^*\cdot\nabla)\tilde{\bm u}
       +\nu\nabla\times\nabla\times\tilde{\bm u}+\nabla p=\bm f^{n+1}.
\label{eq:vms-recovered-momentum}
\end{equation}

These three momentum balances are equivalent at the space-continuous, time-discrete level. The last two balances contain the corrected-velocity acceleration \(\bm a\), whose evaluation in the pressure-increment step involves the unknown current velocity correction. Section~\ref{subsec:vms-local-model} introduces a simplified acceleration choice to address this issue.

\subsection{Multiscale Decomposition and Coarse-Scale Equations}
\label{subsec:vms-coarse-equations}

In this section, we identify the subscale contributions retained in the four RPCS equations before specifying their algebraic models. Sections~\ref{subsec:vms-local-model} and~\ref{subsec:vms-divergence-control} close the velocity- and pressure-subscale contributions, respectively.

Let \(\mathcal T_h\) be a conforming mesh, \(M=L^2(\Omega)\) the pressure space, and \(M_h\subset H^1(\Omega)\subset M\) a continuous Lagrange space of degree \(\ell\ge1\), using \(P_\ell\) or \(Q_\ell\) elements. The homogeneous velocity test space and pressure-increment space are
\begin{equation}
\begin{aligned}
 X&=\{\bm v\in[H^1(\Omega)]^d:\bm v|_{\Gamma_D}=\bm0\},\\
 X_h&=[M_h]^d\cap X,\qquad
 M_{h,0}=\{q_h\in M_h:q_h|_{\Gamma_N}=0\}.
\end{aligned}
\label{eq:vms-spaces}
\end{equation}
The intermediate and corrected velocities belong to \(X_{h,D}^{n+1}=\{\bm v_h\in[M_h]^d:\bm v_h|_{\Gamma_D}=\bm u_{D,h}^{n+1}\}\). For a closed domain, take \(M_{h,0}=M_h\cap L^2_0(\Omega)\), fix the pressure mean, and assume zero net prescribed flux. The \(L^2(\Omega)\) inner product and norm are denoted by \((\cdot,\cdot)\) and \(\|\cdot\|\), respectively. For each element \(K\in\mathcal T_h\), let \((\cdot,\cdot)_K\) denote the \(L^2(K)\) inner product. We define \((\cdot,\cdot)_h=\sum_{K\in\mathcal T_h}(\cdot,\cdot)_K\) and the weighted norm \(\|z\|_\tau^2=(\tau z,z)_h\). A subscript \(h\) on a differential operator denotes elementwise differentiation, while a boundary subscript on an inner product indicates integration over the corresponding boundary.

Form the velocity history \(\bm U_h^n\) and extrapolated advecting velocity \(\bm u_h^*\) from Eq.~\eqref{eq:bdf-history-notation}. Define the discrete BDF accelerations associated with the intermediate and corrected velocities and the provisional pressure by
\begin{equation}
 \widetilde{\bm a}_h=\frac{\gamma_0\tilde{\bm u}_h-\bm U_h^n}{\Delta t},
 \qquad \bm a_h=\frac{\gamma_0\bm u_h-\bm U_h^n}{\Delta t},
 \qquad \pi_h=p_h^n+\phi_h.
\label{eq:vms-predictor-acceleration}
\end{equation}
With \(\beta^*=(\bm u_h^*\cdot\bm n)_-\), the forcing and open-boundary data enter through \(\mathcal L(\bm v_h)=(\bm f^{n+1},\bm v_h)+(\bm h^{n+1},\bm v_h)_{\Gamma_N}\).

Following the VMS framework, we decompose the intermediate velocity and pressure into resolved (coarse) and unresolved (fine) scales. After lifting the prescribed velocity data, let complementary subspaces \(X'\subset X\) and \(M'\subset M\) satisfy \(X=X_h\oplus X'\) and \(M=M_h\oplus M'\). The resulting decompositions are
\begin{equation}
\tilde{\bm u}=\tilde{\bm u}_h+\tilde{\bm u}',
\qquad \tilde{\bm u}'\in X',
\qquad p=p_h+p',
\qquad p'\in M'.
\end{equation}
Here, \(\tilde{\bm u}_h\) and \(p_h\) represent the resolved finite element fields, while \(\tilde{\bm u}'\) and \(p'\) represent the unresolved scales. We model the velocity subscale locally using element bubbles, with \(\tilde{\bm u}'|_K\in[H^1_0(K)]^d\), following the VMS construction \citep{Hughes1995Multiscale,Hughes1998Variational}.

To formulate the convective coupling between the resolved and unresolved velocities, we use the skew convection operator in the weak momentum equation. This choice avoids a spurious interior kinetic-energy contribution when the discrete advecting velocity is not pointwise divergence-free \citep{Charnyi2017Conservation}. For an advecting velocity \(\bm w\), define
\begin{equation}
 (\mathcal C(\bm w)\bm v,\bm z)
 :=((\bm w\cdot\nabla)\bm v,\bm z)
       +\tfrac12((\nabla\cdot\bm w)\bm v,\bm z).
\label{eq:skew-operator}
\end{equation}
Testing this form with \(\bm z=\bm v\) leaves only the convective boundary flux. The momentum residuals used in the algebraic subscale models retain the advective convection term from Section~\ref{subsec:vms-scale-decomposition}, following the distinction discussed by Codina et al.~\citep{Codina2017Variational}. The skew and advective convection expressions coincide when \(\nabla\cdot\bm w=0\).

In the intermediate-velocity momentum equation, the coarse pressure is evaluated at the known value \(p_h^n\). We neglect transient and viscous velocity-subscale feedback \citep{Codina2017Variational,Tezduyar1992Incompressible} and retain the convective coupling \((\mathcal C(\bm u_h^*)\tilde{\bm u}',\bm v_h)_h\). This coupling yields SUPG-type residual stabilization with a skew convection weighting, similar to the scale-splitting treatment of Colom{\'e}s et al.~\citep{Colomes2015Assessment}. The pressure-subscale contribution \(-(p',\nabla\cdot\bm v_h)_h\) provides divergence stabilization through an algebraic model based on the continuity residual, as developed in Section~\ref{subsec:vms-divergence-control} \citep{Olshanskii2009Graddiv}.

Both the pressure-increment and rotational pressure-recovery equations depend on the divergence of the intermediate velocity. Substituting \(\tilde{\bm u}=\tilde{\bm u}_h+\tilde{\bm u}'\) therefore introduces a contribution from \(\nabla\cdot\tilde{\bm u}'\) into each equation. We retain and model these contributions to construct stabilization for the two scalar problems, whose unknowns are \(\phi_h\in M_{h,0}\) and \(p_h\in M_h\), respectively. This treatment is inspired by VMS formulations that model velocity subscales while approximating \(p\approx p_h\) \citep{Codina2017Variational}.

For velocity correction, we project \(\tilde{\bm u}_h-\delta\nabla\phi_h\) onto \(X_{h,D}^{n+1}\). This closure uses the resolved intermediate velocity. Testing the four RPCS steps with \(\bm v_h\in X_h\), \(q_h\in M_{h,0}\), and \(r_h\in M_h\) gives the modeled coarse equations in the order of the algorithm:
\begin{subequations}
\label{eq:vms-weak-problem}
\begin{align}
 &(\widetilde{\bm a}_h,\bm v_h)
 +(\mathcal C(\bm u_h^*)\tilde{\bm u}_h,\bm v_h)
 +\nu(\nabla\tilde{\bm u}_h,\nabla\bm v_h)
 -(p_h^n,\nabla\cdot\bm v_h)-(p',\nabla\cdot\bm v_h)_h\notag\\
 &\hspace{2em}-\tfrac12(\beta^*\tilde{\bm u}_h,\bm v_h)_{\Gamma_N}
 +(\mathcal C(\bm u_h^*)\tilde{\bm u}',\bm v_h)_h
 =\mathcal L(\bm v_h),
\label{eq:vms-coarse-intermediate-velocity}\\
 &\delta(\nabla\phi_h,\nabla q_h)
        =-(\nabla\cdot\tilde{\bm u}_h,q_h)
         -(\nabla\cdot\tilde{\bm u}',q_h)_h,
\label{eq:vms-coarse-pressure-increment}\\
 &(\bm u_h,\bm v_h)=(\tilde{\bm u}_h,\bm v_h)
        -\delta(\nabla\phi_h,\bm v_h),
\label{eq:vms-coarse-velocity-correction}\\
 &(p_h,r_h)=(\pi_h,r_h)-\nu(\nabla\cdot\tilde{\bm u}_h,r_h)
        -\nu(\nabla\cdot\tilde{\bm u}',r_h)_h,
\label{eq:vms-coarse-pressure-recovery}
\end{align}
\end{subequations}
\subsection{Velocity-Subscale Models}
\label{subsec:vms-local-model}

To close the velocity-subscale contributions in Eqs.~\eqref{eq:vms-coarse-intermediate-velocity}, \eqref{eq:vms-coarse-pressure-increment}, and~\eqref{eq:vms-coarse-pressure-recovery}, we construct a momentum residual for each equation from the balances in Section~\ref{subsec:vms-scale-decomposition}. The two pressure-stage balances (Eqs. \eqref{eq:vms-continuous-momentum}, \eqref{eq:vms-recovered-momentum}) involve the corrected-velocity acceleration \(\bm a_h\). To express it in terms of the intermediate velocity, let \(\mathsf P_h:[L^2(\Omega)]^d\to X_h\) be the \(L^2\) projection defined by \((\mathsf P_h\bm z,\bm v_h)=(\bm z,\bm v_h)\) for every \(\bm v_h\in X_h\). Eq.~\eqref{eq:vms-coarse-velocity-correction} gives
\[
 \bm u_h=\tilde{\bm u}_h-\delta\mathsf P_h\nabla\phi_h,
 \qquad
 \bm a_h=\widetilde{\bm a}_h-\mathsf P_h\nabla\phi_h.
\]
In the residual based on Eq.~\eqref{eq:vms-continuous-momentum}, the acceleration correction combines with \(\nabla\pi_h\) to give \(\nabla p_h^n+\nabla\phi_h-\mathsf P_h\nabla\phi_h\). Retaining this expression would place a velocity mass-matrix inverse inside the pressure-increment operator. We therefore replace \(\bm a_h\) by \(\widetilde{\bm a}_h\) in this residual, omitting \(-\mathsf P_h\nabla\phi_h\), so that the increment enters through its gradient alone.

For the rotational pressure-recovery equation, we use the same residual acceleration \(\widetilde{\bm a}_h\), although the corrected velocity is already available. This common choice allows the two pressure-stabilization contributions to be combined in the energy balance of Section~\ref{sec:stability}. This modeling approximation is confined to the subscale residuals and leaves the velocity correction~\eqref{eq:vms-coarse-velocity-correction} unchanged. The temporal accuracy of the resulting formulation, including this residual approximation, is assessed through convergence study in the manufactured-solution tests of Section~\ref{subsec:manufactured-convergence}.

The resulting momentum residuals for Eqs.~\eqref{eq:vms-coarse-intermediate-velocity}, \eqref{eq:vms-coarse-pressure-increment}, and~\eqref{eq:vms-coarse-pressure-recovery}, respectively, are
\begin{subequations}
\label{eq:vms-scalar-residuals}
\begin{align}
 \bm R_m&=\widetilde{\bm a}_h
  +(\bm u_h^*\cdot\nabla)\tilde{\bm u}_h
  -\nu\nabla_h^2\tilde{\bm u}_h+\nabla p_h^n-\bm f^{n+1},
\label{eq:vms-momentum-residual}\\
 \bm R_\phi&=\widetilde{\bm a}_h
  +(\bm u_h^*\cdot\nabla)\tilde{\bm u}_h
  -\nu\nabla_h^2\tilde{\bm u}_h+\nabla\pi_h-\bm f^{n+1},
\label{eq:vms-pressure-increment-residual}\\
 \bm R_p&=\widetilde{\bm a}_h
  +(\bm u_h^*\cdot\nabla)\tilde{\bm u}_h
  +\nu\nabla_h\times\nabla_h\times\tilde{\bm u}_h+\nabla p_h-\bm f^{n+1}.
\label{eq:vms-pressure-recovery-residual}
\end{align}
\end{subequations}
The residual \(\bm R_m\) depends on the current intermediate velocity. After that velocity is computed, \(\bm R_\phi\) and \(\bm R_p\) depend on their respective scalar unknowns through \(\nabla\pi_h=\nabla(p_h^n+\phi_h)\) and \(\nabla p_h\).

We integrate the retained bubble contributions by parts within each element, transferring derivatives to the coarse test functions. The zero bubble traces eliminate the boundary terms in this integration. Following Hughes's construction, the local fine-scale response is formally represented through an element Green's operator \citep{Hughes1995Multiscale,Hughes1998Variational}. We approximate its contribution to the coarse-test integrals using the stage-specific algebraic models
\begin{equation}
 \tilde{\bm u}'_m \approx -\tau_m\bm R_m,
 \qquad \tilde{\bm u}'_\phi \approx -\tau_m\bm R_\phi,
 \qquad \tilde{\bm u}'_p \approx -\tau_m\bm R_p,
\label{eq:vms-algebraic-subscales}
\end{equation}
where \(\tau_{m}>0\) summarizes the local response. The algebraic expressions are substituted into the coarse-test integrals, giving the following contributions on the left-hand sides of the three equations:
\begin{equation}
\begin{aligned}
 -(\tilde{\bm u}'_m,\mathcal C(\bm u_h^*)\bm v_h)_h
 &\approx(\tau_m\bm R_m,\mathcal C(\bm u_h^*)\bm v_h)_h,\\
 -(\tilde{\bm u}'_\phi,\nabla q_h)_h
 &\approx(\tau_m\bm R_\phi,\nabla q_h)_h,\\
 -\nu(\tilde{\bm u}'_p,\nabla r_h)_h
 &\approx\nu(\tau_m\bm R_p,\nabla r_h)_h.
\end{aligned}
\label{eq:vms-stage-feedback}
\end{equation}
The first is the convective stabilization term introduced above. Together with the continuity-residual model for \(p'\) in Section~\ref{subsec:vms-divergence-control}, it stabilizes the intermediate-velocity equation and controls its divergence. The other two terms pair the appropriate momentum residual with a scalar test-function gradient, supplying stabilization to the pressure-increment and rotational pressure-recovery equations. The resulting scheme admits a conditional bound on the pressure gradient for the reduced model analyzed in Section~\ref{sec:stability}.

For all three models, we use the metric-based momentum scale in element $K$ \citep{Bazilevs2007Variational}
\begin{equation}
 \tau_{m,K}=\left[\left(\frac{\gamma_0}{\Delta t}\right)^2
       +\bm u_{h,K}^*\cdot\bm G_K\bm u_{h,K}^*
       +C_I\nu^2\bm G_K:\bm G_K\right]^{-1/2},
       \qquad C_I>0.
\label{eq:vms-parameters}
\end{equation}
Here \(\bm u_{h,K}^*\) is a representative element velocity, and \((G_K)_{ij}=\sum_{a=1}^d(\partial\xi_a/\partial x_i)(\partial\xi_a/\partial x_j)\), where \(\xi_a\) are reference coordinates. \(\bm G_K:\bm G_K=\sum_{i,j}(G_K)_{ij}^2\) scales as inverse length to the fourth power and is invariant under orthogonal changes of reference coordinates. The time-step term ensures \(0<\tau_{m,K}\le\delta \le \Delta t\).

\subsection{Pressure-Subscale Model}
\label{subsec:vms-divergence-control}
\label{subsec:vms-velocity-stages}

The pressure-subscale term in Eq.~\eqref{eq:vms-coarse-intermediate-velocity} provides divergence control through a continuity-residual model. This follows the VMS interpretation of grad-div stabilization as an algebraic pressure-subscale contribution \citep{Olshanskii2009Graddiv,tenEikelderAkkerman2018CorrectEnergyII}. For RPCS, the relevant continuity relation includes the pressure increment. Taking the divergence of the space-continuous velocity correction gives
\begin{equation}
 \nabla\cdot\bm u=\nabla\cdot\tilde{\bm u}-\delta\nabla^2\phi=0.
\label{eq:vms-continuity-residual}
\end{equation}
Evaluating this relation on the finite-element fields defines the elementwise residual
\begin{equation}
 R_c=\nabla\cdot\tilde{\bm u}_h-\delta\nabla_h^2\phi_h.
\label{eq:vms-discrete-continuity-residual}
\end{equation}
We represent the pressure subscale by the algebraic field \(p'=-\tau_cR_c\), with a nonnegative elementwise coefficient \(\tau_c\). Its contribution to the intermediate-velocity equation is
\begin{equation}
 -(p',\nabla\cdot\bm v_h)_h
 =(\tau_cR_c,\nabla\cdot\bm v_h)_h.
 \label{eq:vms-continuity-subscale}
\end{equation}

For affine $P_1$ elements, the elementwise pressure Laplacian vanishes, and the pressure-subscale model reduces to classical grad--div stabilization \citep{Olshanskii2009Graddiv}. When the elementwise pressure Laplacian is nonzero, the pressure-increment contribution can be evaluated using extrapolation from previous time levels, with \(\phi_h^*=\phi_h^n\) for BDF1 and \(\phi_h^*=2\phi_h^n-\phi_h^{n-1}\) for BDF2. This preserves the sequential structure of the four-step scheme.

To set the strength of the pressure-subscale contribution, we use the metric-based continuity scale \citep{tenEikelderAkkerman2018CorrectEnergyII,BazilevsAkkerman2010TaylorCouette}, with a nonnegative element factor \(\alpha_{c,K}\):
\begin{equation}
 \tau_{c,K}=\frac{\alpha_{c,K}}
       {\tau_{m,K}(\bm G_K:\bm G_K)^{1/2}},
 \qquad\alpha_{c,K}\ge0.
\label{eq:vms-continuity-parameter}
\end{equation}
\(\alpha_{c,K}\) is dimensionless and \(\tau_{c,K}\) has the units of kinematic viscosity. The factor \(\alpha_{c,K}\) may vary by element, and both stabilization coefficients are fixed during each step.

\subsection{Fully Discrete RPCS-VMS Scheme}
\label{subsec:vms-implementation}

We now incorporate the velocity- and pressure-subscale models into the four-step RPCS framework to obtain the fully discrete RPCS-VMS scheme. At each step, form the BDF data from Eq.~\eqref{eq:bdf-history-notation}, the coefficients from Eqs.~\eqref{eq:vms-parameters} and~\eqref{eq:vms-continuity-parameter}, and the extrapolated pressure increment from Section~\ref{subsec:vms-divergence-control} if needed. Keep these data and \(\beta^*\) fixed during the four solves. The tests are \(\bm v_h\in X_h\), \(q_h\in M_{h,0}\), and \(r_h\in M_h\).

\noindent\textit{Step 1: Intermediate velocity.}
Given the velocity history and \(p_h^n\), solve for the intermediate velocity \(\tilde{\bm u}_h\in X_{h,D}^{n+1}\):
\begin{equation}
\begin{aligned}
 &\left(\frac{\gamma_0\tilde{\bm u}_h-\bm U_h^n}{\Delta t},\bm v_h\right)
 +(\mathcal C(\bm u_h^*)\tilde{\bm u}_h,\bm v_h)
 +\nu(\nabla\tilde{\bm u}_h,\nabla\bm v_h)
 -(p_h^n,\nabla\cdot\bm v_h)\\
 &\quad-\tfrac12(\beta^*\tilde{\bm u}_h,\bm v_h)_{\Gamma_N}
 +(\tau_m\bm R_m,\mathcal C(\bm u_h^*)\bm v_h)_h\\
 &\quad+\left(\tau_c\left[\nabla\cdot\tilde{\bm u}_h
       -\frac{\Delta t}{\gamma_0}\nabla_h^2\phi_h^*\right],
                      \nabla\cdot\bm v_h\right)_h
 =(\bm f^{n+1},\bm v_h)+(\bm h^{n+1},\bm v_h)_{\Gamma_N}.
\end{aligned}
\label{eq:vms-intermediate-velocity}
\end{equation}

\noindent\textit{Step 2: Pressure increment.}
Using \(\tilde{\bm u}_h\) from Step~1, evaluate \(\bm R_m\). With \(\bm R_\phi=\bm R_m+\nabla\phi_h\), solve for \(\phi_h\in M_{h,0}\):
\begin{equation}
\begin{aligned}
 \frac{\Delta t}{\gamma_0}(\nabla\phi_h,\nabla q_h)
 +(\tau_m\nabla\phi_h,\nabla q_h)_h
 =-(\nabla\cdot\tilde{\bm u}_h,q_h)
   -(\tau_m\bm R_m,\nabla q_h)_h.
\end{aligned}
\label{eq:vms-pressure-increment}
\end{equation}
\noindent\textit{Step 3: Velocity correction.}
Using \(\phi_h\) from Step~2, solve for the corrected velocity \(\bm u_h\in X_{h,D}^{n+1}\):
\begin{equation}
 (\bm u_h,\bm v_h)=(\tilde{\bm u}_h,\bm v_h)
 -\frac{\Delta t}{\gamma_0}(\nabla\phi_h,\bm v_h).
\label{eq:vms-velocity-correction}
\end{equation}

\noindent\textit{Step 4: Rotational pressure recovery.}
Using \(\tilde{\bm u}_h\) and \(\phi_h\), solve for \(p_h\in M_h\):
\begin{equation}
\begin{aligned}
 &(p_h,r_h)+\nu(\tau_m\nabla p_h,\nabla r_h)_h
 =(p_h^n+\phi_h,r_h)-\nu(\nabla\cdot\tilde{\bm u}_h,r_h)\\
 &\quad-\nu\Bigl(\tau_m\Bigl[
 \frac{\gamma_0\tilde{\bm u}_h-\bm U_h^n}{\Delta t}
 +(\bm u_h^*\cdot\nabla)\tilde{\bm u}_h
 +\nu\nabla_h\times\nabla_h\times\tilde{\bm u}_h
 -\bm f^{n+1}\Bigr],\nabla r_h\Bigr)_h.
\end{aligned}
\label{eq:vms-pressure-recovery}
\end{equation}

\section{Stability Analysis of the RPCS-VMS Formulation}
\label{sec:stability}

To isolate the interaction between pressure stabilization and rotational recovery, we analyze a reduced BDF1 version of the four-step scheme in Section~\ref{subsec:vms-implementation}. The body force and prescribed velocity and pseudo-traction data are zero. Related analyses have examined pressure segregation without convective stabilization \citep{Codina2006On} and rotational pressure correction for convection-free problems \citep{Guermond2006Error,LeeSalgado2016Stability}. Following these studies, the intermediate-velocity equation retains skew-symmetric convection and the backflow boundary term, but omits SUPG stabilization. The two pressure equations retain their stabilization terms, with convection omitted from the associated momentum residuals.

\subsection{Discrete energy structure}
\label{subsec:stokes-energy-setting}

We use continuous \(P_1/P_1\) fields on a fixed affine simplicial mesh, and their elementwise second derivatives vanish. The mesh is connected through shared faces and \(\Gamma_D\) has positive surface measure. The initial fields belong to \(X_h\times M_h\), with a prescribed pressure mean for a closed domain. At each step the known advecting field is \(\bm w_h^{n+1}=\bm u_h^*\), with \(\beta^*=(\bm w_h^{n+1}\cdot\bm n)_-\). The skew convection form and backflow term together give a nonnegative outflow contribution: for \(\bm v_h\in X_h\),
\begin{equation}
 2(\mathcal C(\bm w_h^{n+1})\bm v_h,\bm v_h)
 -(\beta^*\bm v_h,\bm v_h)_{\Gamma_N}
 =\int_{\Gamma_N}(\bm w_h^{n+1}\cdot\bm n)_+
                         |\bm v_h|^2\,\mathrm{d}s\ge0,
 \qquad s_+=\max(s,0).
\label{eq:stability-oseen-flux}
\end{equation}
For the stabilization coefficients, define the extrema over the mesh by
\[
 \tau_m^{\min}=\min_K\tau_{m,K},\qquad
 \tau_m^{\max}=\max_K\tau_{m,K},\qquad
 \tau_c^{\min}=\min_K\tau_{c,K}.
\]
For reasons that will become clear in the energy and pressure estimates below,
assume that the element coefficients are constant in time and satisfy
\begin{equation}
 0<\tau_m^{\min}\le\tau_m^{\max}\le\Delta t,
 \qquad
 \kappa_c:=2\tau_c^{\min}
          -\left(1-\frac2d\right)\nu\ge0.
 \label{eq:stability-assumptions}
\end{equation}
The condition on \(\tau_c\) requires \(\tau_{c,K}\ge0\) in two dimensions and \(\tau_{c,K}\ge\nu/6\) in three dimensions.

Recall from Section~\ref{subsec:vms-coarse-equations} that \(\pi_h^{n+1}=p_h^n+\phi_h^{n+1}\) is the provisional pressure. The elementwise viscous terms in Eq.~\eqref{eq:vms-scalar-residuals} vanish for \(P_1\) velocities. With zero body force and convection omitted from these residuals, the momentum residuals used in the pressure-increment and pressure-recovery equations reduce to \(\bm R_\phi^{n+1}=\widetilde{\bm a}_h^{n+1}+\nabla\pi_h^{n+1}\) and \(\bm R_p^{n+1}=\widetilde{\bm a}_h^{n+1}+\nabla p_h^{n+1}\), where \(\widetilde{\bm a}_h^{n+1}=(\widetilde{\bm u}_h^{n+1}-\bm u_h^n)/\Delta t\) is the BDF1 intermediate-velocity acceleration. The four-step equations used in the analysis are listed in Appendix~\ref{app:energy-identity}.

To convert the pressure-increment contribution into an energy difference, introduce the accumulated increment \(\psi_h^n\in M_{h,0}\):
\begin{equation}
 \psi_h^0=0,\qquad
 \psi_h^{n+1}=\psi_h^n+\phi_h^{n+1}.
\label{eq:stability-auxiliary-def}
\end{equation}
Since \(\phi_h^{n+1}=\psi_h^{n+1}-\psi_h^n\), testing the pressure increment equation with \(\psi_h^{n+1}\) produces a difference of squared gradient norms.

To account for rotational recovery in the same balance, define \(\eta_h^n=\psi_h^n-p_h^n\in M_h\). Its change is exactly the difference between the provisional and final pressures:
\[
 \delta\eta_h^{n+1}:=\eta_h^{n+1}-\eta_h^n
 =\pi_h^{n+1}-p_h^{n+1}.
\]
The residual definitions in Section~\ref{subsec:vms-local-model} imply \(\bm R_p-\bm R_\phi=\nabla_h(p_h-\pi_h+\nu\nabla\cdot\tilde{\bm u}_h)\). On the present \(P_1\) mesh, \(\nabla_h(\nabla\cdot\tilde{\bm u}_h)=\bm0\), so
\[
 \bm R_p^{n+1}-\bm R_\phi^{n+1}
 =-\nabla\delta\eta_h^{n+1}.
\]
This relation connects the two pressure-stabilization contributions when completing squares in the energy balance, as detailed in Appendix~\ref{app:energy-identity}. The sequences \(\psi_h^n\) and \(\eta_h^n\) are introduced only for the analysis and require no additional pressure solves.

The relation \(\pi_h^{n+1}=\psi_h^{n+1}-\eta_h^n\) connects the increment and recovery tests. Testing the pressure-recovery relation with \(\eta_h^n\) completes the conversion of pressure work into differences of squared norms. These squared norms form the pressure terms in the augmented energy \(E^n\). The viscous and divergence contributions are grouped in \(\mathscr D_h\):
\begin{equation}
\begin{aligned}
 E^n&=\|\bm u_h^n\|^2+(\Delta t)^2\|\nabla\psi_h^n\|^2
       +\frac{\Delta t}{\nu}\|\eta_h^n\|^2
       +\Delta t\|\nabla\eta_h^n\|_{\tau_m}^2,\\
 \mathscr D_h(\bm v_h)&=
       2\nu\|\nabla\bm v_h\|^2
       +2\|\nabla\cdot\bm v_h\|_{\tau_c}^2
       -\nu\|\nabla\cdot\bm v_h\|^2.
\end{aligned}
\label{eq:stability-energy-def}
\end{equation}
Under Eq.~\eqref{eq:stability-assumptions}, \(\mathscr D_h\) is positive definite on \(X_h\), including at \(\kappa_c=0\) (Appendix~\ref{app:energy-identity}). This property also provides the divergence control needed for the pressure estimate.

\subsection{Energy decay}
\label{subsec:energy-estimate}

Testing the intermediate-velocity equation and applying the projected correction gives an energy balance for the corrected velocity. The pressure-increment and pressure-recovery equations then express pressure work as changes in the auxiliary energy terms. Since \(\tau_m\) is constant in time, the weighted energy differences telescope when summed over successive steps.

To express the remaining terms in the energy balance, write \(\mathsf Q_h=I-\mathsf P_h\), the complement of the velocity projection, where \(I\) is the identity on \([L^2(\Omega)]^d\). We collect the contributions associated with the time increment, projected correction, viscous and stabilization terms, and outflow as
\begin{equation}
\begin{aligned}
 W_n&=\sum_K\left(1-\frac{\tau_{m,K}}{\Delta t}\right)
       \|\widetilde{\bm u}_h^{n+1}-\bm u_h^n\|_{L^2(K)}^2,\\
 J_n&=(\Delta t)^2\|\mathsf Q_h\nabla\phi_h^{n+1}\|^2,\\
 F_n&=\mathscr D_h(\widetilde{\bm u}_h^{n+1})
                  +\|\nabla\pi_h^{n+1}\|_{\tau_m}^2,\\
 B_n&=\int_{\Gamma_N}(\bm w_h^{n+1}\cdot\bm n)_+
                  |\widetilde{\bm u}_h^{n+1}|^2\,\mathrm{d}s,
\end{aligned}
\label{eq:stability-dissipation-def}
\end{equation}
while pressure recovery contributes
\begin{equation}
 \mathcal S_n=\Delta t\|\bm R_p^{n+1}\|_{\tau_m}^2
  +\frac{\Delta t}{\nu}
       \|\delta\eta_h^{n+1}-\nu \nabla\cdot\widetilde{\bm u}_h^{n+1}\|^2.
\label{eq:stability-recovery-dissipation}
\end{equation}

To complete the balance, we take \(r_h=\delta\eta_h^{n+1}\) in the pressure-recovery relation, Eq.~\eqref{eq:stability-eta-relation}, and combine the residual and divergence contributions by completing squares. This produces \(\mathcal S_n\) and the pressure-gradient term in \(F_n\), together with a negative squared norm of the intermediate-velocity acceleration. That negative term combines with the squared velocity increment from the test of the intermediate-velocity equation to give \(W_n\). Appendix~\ref{app:energy-identity} provides the full derivation.

Under the assumptions of Section~\ref{subsec:stokes-energy-setting}, the solution of the reduced four-step scheme satisfies the discrete energy-dissipation identity
\begin{equation}
 E^{n+1}-E^n+W_n+J_n+\Delta t F_n+\mathcal S_n+\Delta t B_n=0.
\label{eq:stability-closed-step}
\end{equation}
Hence, for every \(N\ge1\),
\begin{equation}
 E^N+\sum_{n=0}^{N-1}(W_n+J_n+\Delta t F_n+\mathcal S_n+\Delta t B_n)=E^0,
 \qquad E^{n+1}\le E^n.
\label{eq:stability-global-energy}
\end{equation}

The upper bound \(\tau_m^{\max}\le\Delta t\) ensures \(W_n\ge0\), and the condition \(\kappa_c\ge0\) ensures \(\mathscr D_h\ge0\). The remaining contributions are squared norms or the nonnegative outflow flux in Eq.~\eqref{eq:stability-oseen-flux}. Thus all dissipation terms in Eq.~\eqref{eq:stability-closed-step} are nonnegative, giving energy decay for both closed and open boundaries.

\subsection{Pressure-gradient bounds}
\label{subsec:pressure-stability}

The energy identity controls the gradient of the provisional pressure through \(\|\nabla\pi_h^{n+1}\|_{\tau_m}^2\) in \(F_n\). To obtain a bound for the gradient of the updated pressure \(p_h^{n+1}\), we use \(p_h^{n+1}=\pi_h^{n+1}-\delta\eta_h^{n+1}\). The second term in \(\mathcal S_n\) controls \(\delta\eta_h^{n+1}-\nu\nabla\cdot\widetilde{\bm u}_h^{n+1}\), and \(\mathscr D_h\) controls the divergence. Together they bound the change \(\delta\eta_h^{n+1}\). A finite-element inverse estimate then bounds its gradient. Summing these contributions in time and using the energy identity gives, for \(\tau_m^{\min}>0\), a finite constant \(C_{p,h}\), independent of the number of steps, such that
\begin{equation}
 \Delta t\sum_{n=0}^{N-1}\|\nabla p_h^{n+1}\|^2
 \le\frac{C_{p,h}}{\tau_m^{\min}}E^0.
\label{eq:stability-pressure-gradient}
\end{equation}
The positive lower bound \(\tau_m^{\min}\) converts the weighted pressure-gradient control into this unweighted estimate. The constant \(C_{p,h}\) depends on the mesh and frozen coefficients and is specified in Appendix~\ref{app:energy-pressure}. On a fixed mesh, Eq.~\eqref{eq:vms-parameters} gives \(\tau_m^{\min}\) proportional to \(\Delta t\) as \(\Delta t\to0\), provided the advecting velocity is bounded in \(L^\infty(\Omega)\) uniformly in the time level and \(\Delta t\). The explicit factor \(1/\tau_m^{\min}\) therefore grows as \((\Delta t)^{-1}\).

For comparison, we remove the \(\tau_m\)-terms and prescribe \(\tau_c\) independently so that \(\kappa_c\ge0\). Energy decay still holds, but the pressure-gradient dissipation is absent. With \(E_0^0=\|\bm u_h^0\|^2+(\Delta t/\nu)\|p_h^0\|^2\) and \(T=N\Delta t\), Appendix~\ref{app:energy-unstabilized} gives
\begin{equation}
 \Delta t\sum_{n=0}^{N-1}\|\nabla p_h^{n+1}\|^2
 \le\frac{2T}{(\Delta t)^2}
       \max\{1,\nu\Delta t C_{\nabla,h}^2\}\,E_0^0.
\label{eq:stability-pressure-without-vms}
\end{equation}
Here \(C_{\nabla,h}\) is a mesh-dependent constant satisfying \(\|\nabla r_h\|\le C_{\nabla,h}\|r_h\|\) for every \(r_h\in M_h\). At fixed mesh and final time, the factor on the right grows as \((\Delta t)^{-2}\), compared with the explicit \((\Delta t)^{-1}\) factor in the stabilized estimate when \(\tau_m^{\min}\) is proportional to \(\Delta t\). Stabilization thus supplies pressure-gradient dissipation and improves the time-step dependence of the bound.

As the time step decreases, the estimate without momentum-residual stabilization provides progressively weaker pressure-gradient control and does not rule out pressure oscillations. This behavior is consistent with the small-time-step pressure instability discussed by Codina \citep{Codina2001Pressure}. In particular, a pressure mode orthogonal to all discrete velocity divergences can persist unchanged despite the nonincrease of energy. Appendix~\ref{app:energy-unstabilized} gives this construction.

\suppressfloats[t]

\section{Numerical Results}
\label{sec:results}

In this section, we present the numerical results of the RPCS-VMS method. Manufactured solutions are used to examine spatial and temporal convergence, near-boundary pressure errors, and pressure stability at small time steps. The branched-pipe benchmark assesses flow partitioning between two open outlets, while two-dimensional (2D) and three-dimensional (3D) cylinder benchmarks evaluate transient forces and pressure predictions against FEATFLOW reference data \citep{FeatFlowDFG2D3,FeatFlowDFG3DReference}. The 3D cylinder case also serves to assess computational efficiency relative to a monolithic VMS formulation.

The computations use FEniCSx 0.10.0 \citep{Baratta2025DOLFINx} and PETSc 3.24.3 \citep{PETScUsersManual2026}. The stabilized equal-order discretizations use $C_I=36$ in the momentum scale of Eq.~\eqref{eq:vms-parameters}, following Brunelli et al.~\citep{Brunelli2026Separation}. To meet the divergence-penalty threshold in Eq.~\eqref{eq:stability-assumptions}, the divergence parameter is set elementwise to
\begin{equation}
 \alpha_{c,K}=\left(1-\frac2d\right)\frac{\nu}{2}\,
       \tau_{m,K}(\bm G_K:\bm G_K)^{1/2}.
\label{eq:numerical-alpha-threshold}
\end{equation}
This leads to $\tau_c=0$ in 2D and $\tau_c=\nu/6$ in 3D. Except for the explicitly modified variants in Section~\ref{subsec:pressure-stage-ablation}, all RPCS-VMS computations use the full formulation of Section~\ref{subsec:vms-implementation}.

The intermediate-velocity system is solved using the generalized minimal residual method (GMRES), preconditioned with BoomerAMG, an implementation of algebraic multigrid (AMG). The pressure-increment and pressure-recovery systems are solved using the conjugate gradient method (CG), also preconditioned with BoomerAMG, while the velocity correction system is solved using CG with Jacobi preconditioning. All four solves use a relative tolerance of $10^{-8}$, an absolute tolerance of $10^{-10}$, and a maximum of 1000 iterations. The GMRES restart is set to 100. Table~\ref{tab:numerical_reproducibility} summarizes the meshes, unknown counts, and time steps for the physical cases.

\begin{table}[H]
\centering
\small
\begin{tabular}{lrrrr}
\toprule
Case & Vertices/cells & Element pair & No. of unknowns & $\Delta t,\ T$ \\
\midrule
Branched pipe & 8,086/15,293 & $P_1/P_1$ & 16,172/8,086 & $5\times10^{-4}$,\ 50 \\
2D cylinder & 33,742/33,268 & $Q_1/Q_1$ & 67,484/33,742 & $6.25\times10^{-4}$,\ 8 \\
3D cylinder & 410,800/385,920 & $Q_1/Q_1$ & 1,232,400/410,800 & $5\times10^{-3}$,\ 8 \\
\bottomrule
\end{tabular}
\caption{Mesh and time-integration settings for the physical benchmarks. Unknown counts are listed as velocity/pressure, and $T$ is the final simulation time.}
\label{tab:numerical_reproducibility}
\end{table}

\subsection{Manufactured-Solution Tests}

The convergence and pressure tests use a common manufactured Navier-Stokes solution on the square domain \(\Omega=[-0.5,0.5]^2\). The right edge, \(x=0.5\), carries the open-boundary condition, while the exact velocity is prescribed on the other three edges. Following the spatial fields of Liu \citep{Liu2009Open}, set
\begin{equation}
 \bm u_s=
 \begin{pmatrix}
 \sin(\upi y)\cos^2(\upi x/2)\\
 -\sin(\upi x)\cos^2(\upi y/2)
 \end{pmatrix},
 \qquad p_s=\cos(\upi x/2)\sin(\upi y/2),
 \qquad
 (\bm u,p)=A(t)(\bm u_s,p_s).
\label{eq:manufactured-family}
\end{equation}
The spatial velocity field is divergence-free. We use \(\nu=1\), with \(A=1\) for the steady tests. The unsteady tests use \(A(t)=\cos(2\omega t)\) with \(\omega=\upi\) and end at $T=1$, except the near-boundary pressure comparison at $T=0.5$. The body force and open-boundary data are obtained by substituting these fields into Eqs.~\eqref{eq:strong-momentum} and~\eqref{eq:obc}, respectively.

\subsubsection{Spatial and Temporal Convergence}
\label{subsec:manufactured-convergence}

\begin{figure}[htbp]
    \centering
    \includegraphics[width=\textwidth]{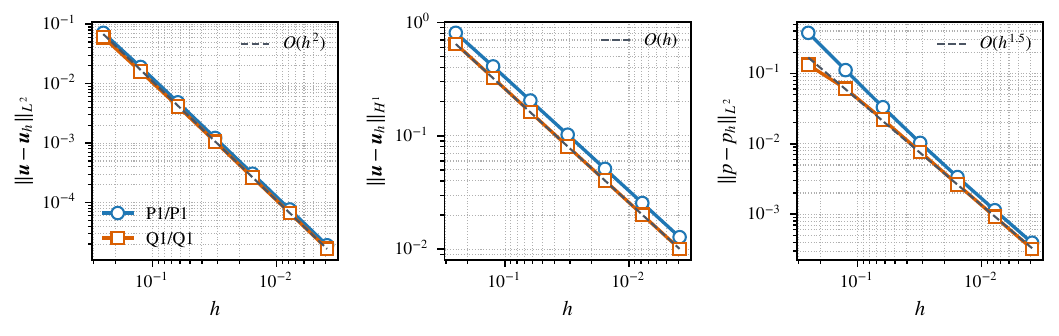}
    \caption{Spatial convergence of RPCS-VMS for the steady manufactured solution with $P_1/P_1$ and $Q_1/Q_1$ elements. Reference slopes indicate the convergence orders.}
    \label{fig:conv_dx}
\end{figure}

We first examine spatial convergence against the steady solution, using BDF1 pseudo-time stepping with $\Delta t=10^{-2}$ to approach steady state. Simulations are performed on $P_1/P_1$ and $Q_1/Q_1$ meshes constructed from $n\times n$ grids, with $n=4,8,\ldots,256$. Fig.~\ref{fig:conv_dx} shows second-order velocity convergence in $L^2$, first-order convergence in the full $H^1$ norm, and approximately order-$1.5$ pressure convergence for both element types.

\begin{figure}[htbp]
    \centering
    \includegraphics[width=\textwidth]{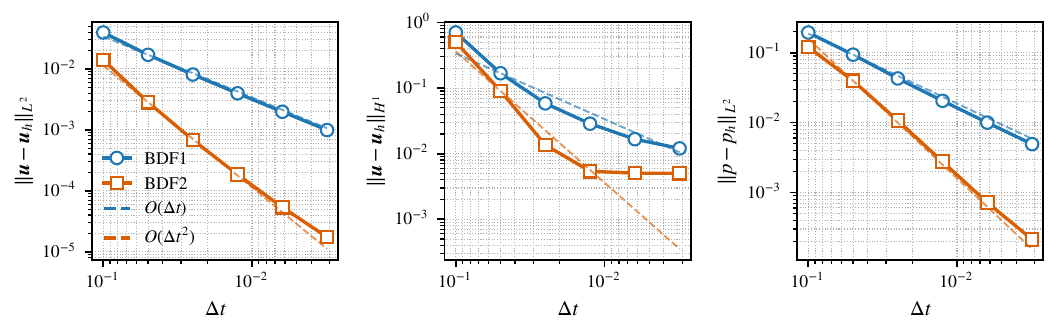}
    \caption{Temporal convergence of RPCS-VMS for the unsteady manufactured solution with $Q_1/Q_1$ elements. BDF1 uses $n=256$ and BDF2 uses $n=512$. Errors are evaluated at $T=1$.}
    \label{fig:conv_dt}
\end{figure}

We next examine temporal convergence against the unsteady reference solution using $Q_1/Q_1$ elements and successively halved time steps ranging from $0.1$ to $0.003125$. Fig.~\ref{fig:conv_dt} shows first-order convergence in the velocity and pressure $L^2$ errors for BDF1 and second-order convergence for BDF2. As the time step decreases, spatial discretization errors become significant and limit further reduction in errors, with the velocity $H^1$ error reaching this limit earlier.

The observed spatial velocity convergence rates are consistent with standard error estimates for stabilized low-order finite element approximations of smooth solutions \citep{FrancaHughesStenberg1993Stabilized}. The pressure $L^2$ convergence rate of approximately $1.5$ is consistent with that previously reported for a stabilized $Q_1/Q_1$ monolithic discretization of Stokes flow \citep{Hughes1986New}.

\subsubsection{Near-Boundary Pressure Errors}
\label{subsec:boundary-pressure-effects}

To investigate the spatial distribution of pressure errors, we compare three formulations: RPCS-VMS, the non-rotational formulation, and Monolithic-VMS. All three use BDF2 with $\Delta t=10^{-3}$ on the same $P_1/P_1$ mesh constructed from a $64\times64$ uniform grid and are evaluated at $T=0.5$. The half-period endpoint reveals the boundary-pressure error before the exact pressure returns to its initial value at $T=1$. The non-rotational formulation retains the pressure-increment stabilization but replaces the rotational pressure-recovery equation with $p_h^{n+1}=p_h^n+\phi_h^{n+1}$. Monolithic-VMS denotes the monolithic equal-order formulation with SUPG/PSPG stabilization and the grad-div term governed by $\tau_c$ \citep{Tezduyar1992Incompressible}. All three formulations use the same $\tau_m$ model and solver tolerances.

Let the boundary element strip and its interior complement be
\begin{equation}
    \Omega_h^b=\bigcup\{K\in\mathcal T_h:K\text{ has a facet on }\partial\Omega\},
    \qquad
    \Omega_h^i=\Omega\setminus\Omega_h^b.
\end{equation}

\begin{figure}[htbp]
    \centering
    \includegraphics[width=\textwidth]{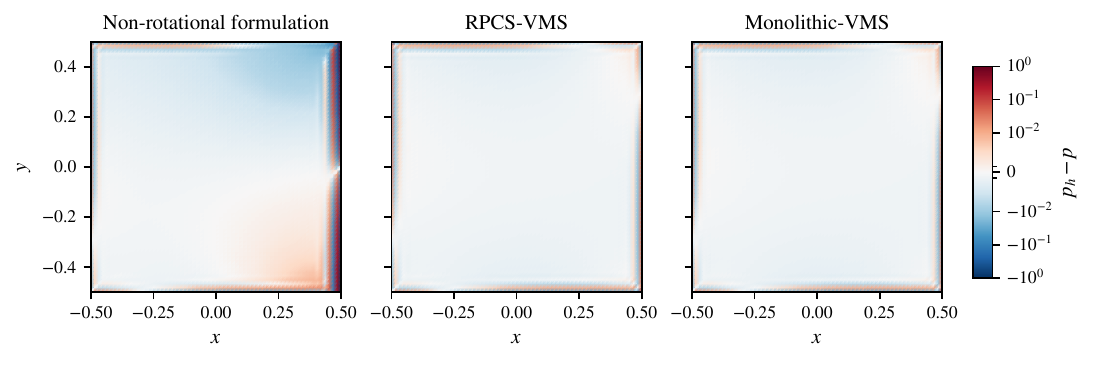}
    \caption{Unsteady manufactured-solution pressure errors $p_h-p$ at $T=0.5$ for the non-rotational formulation, RPCS-VMS, and Monolithic-VMS. The common symmetric logarithmic color scale is linear for $|p_h-p|\le0.005$.}
    \label{fig:boundary_pressure_comparison}
\end{figure}

Table~\ref{tab:boundary_pressure_comparison} reports $L^2$ errors over the full domain, boundary strip, and interior, with $\|\cdot\|_D$ denoting the $L^2(D)$ norm. As shown in Fig. \ref{fig:boundary_pressure_comparison}, the pressure error is concentrated near the open boundary for non-rotational formulation, because Eq. \eqref{eq:rpcs-bc-phi-open} imposed on the pressure increment keeps the non-rotational pressure at its initial value along this boundary. Relative to this formulation, the stabilized rotational recovery treatment reduces the boundary-strip pressure error from $3.4536\times10^{-2}$ to $2.6738\times10^{-3}$, a factor of 12.92, and the global pressure error by a factor of 12.47. RPCS-VMS and Monolithic-VMS errors agree to within $0.4\%$, with nearly identical boundary-adjacent patterns in Fig.~\ref{fig:boundary_pressure_comparison}. Localized stabilization-induced pressure perturbations remain in both formulations.

\par\addvspace{10pt}
\noindent\begin{minipage}{\linewidth}
\begin{center}
\begin{tabular}{lcccc}
\toprule
Method & $\|p-p_h\|_{\Omega}$ & $\|p-p_h\|_{\Omega_h^b}$ & $\|p-p_h\|_{\Omega_h^i}$ & $\|\bm u-\bm u_h\|_{\Omega}$ \\
\midrule
Non-rotational formulation & $4.0855\times10^{-2}$ & $3.4536\times10^{-2}$ & $2.1826\times10^{-2}$ & $6.5958\times10^{-4}$ \\
RPCS-VMS & $3.2775\times10^{-3}$ & $2.6738\times10^{-3}$ & $1.8955\times10^{-3}$ & $3.0185\times10^{-4}$ \\
Monolithic-VMS & $3.2725\times10^{-3}$ & $2.6703\times10^{-3}$ & $1.8917\times10^{-3}$ & $3.0077\times10^{-4}$ \\
\bottomrule
\end{tabular}
\end{center}
\refstepcounter{table}
\label{tab:boundary_pressure_comparison}
{\small
\noindent\textbf{Table~\thetable}\par
\noindent Unsteady manufactured-solution $L^2$ errors at $T=0.5$ with BDF2, $\Delta t=10^{-3}$, and $n=64$. All methods use $P_1/P_1$ elements. Pressure errors are calculated over the full domain $\Omega$, boundary strip $\Omega_h^b$, and interior $\Omega_h^i$. The velocity error is evaluated over the full domain.\par
}
\end{minipage}\par
\par\addvspace{10pt}

\subsubsection{Stability at Small Time Steps}
\label{subsec:manufactured-stability}

\begin{figure}[!htbp]
    \centering
    \includegraphics[width=\textwidth]{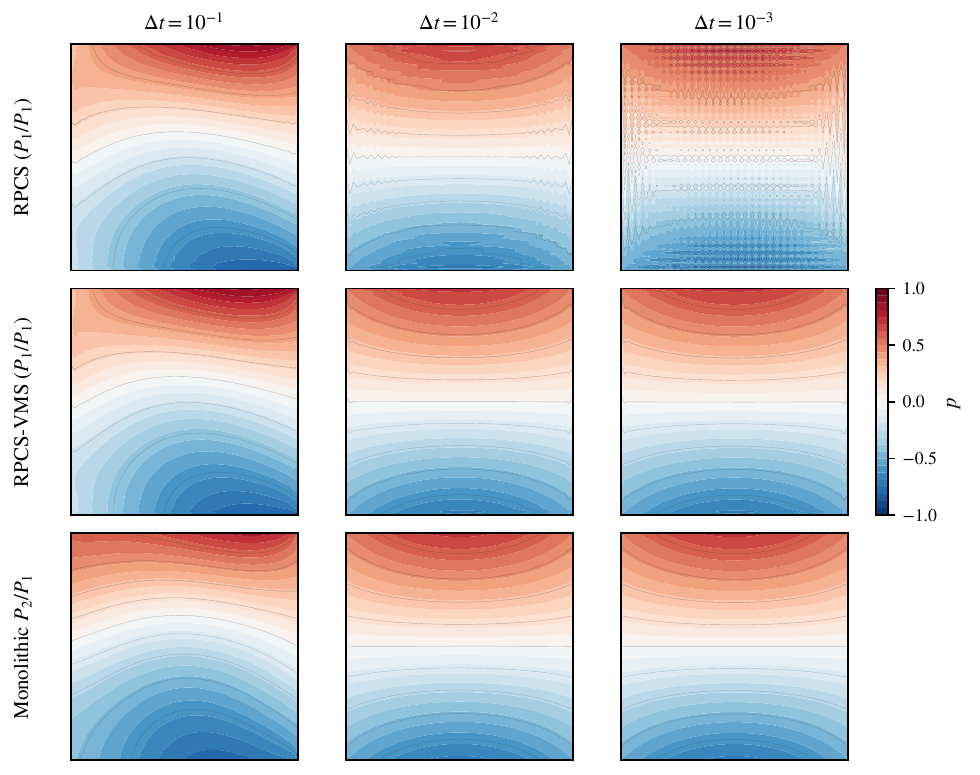}
    \caption{Pressure fields at $T=1$ in the small-time-step manufactured-solution test. All panels use the same color scale.}
    \label{fig:pressure_matrix}
\end{figure}

We test small-time-step pressure stability using BDF2 and the unsteady manufactured solution on a $64\times64$ grid at $T=1$. Fig.~\ref{fig:pressure_matrix} compares unstabilized RPCS and RPCS-VMS with $P_1/P_1$ elements against an unstabilized monolithic $P_2/P_1$ reference at $\Delta t=10^{-1},10^{-2},10^{-3}$. Decreasing $\Delta t$ produces mesh-scale pressure oscillations in RPCS, consistent with Codina's analysis \citep{Codina2001Pressure}, while the RPCS-VMS pressure stays close to the monolithic reference. The reduced-model estimate in Section~\ref{subsec:pressure-stability} identifies the additional pressure-gradient control supplied by the $\tau_m$ terms, consistent with the smoother pressure fields observed here.

\subsubsection{Effects of Stabilization in the Two Pressure Equations}
\label{subsec:pressure-stage-ablation}

To examine the contribution of each pressure stabilization term, we compare the full RPCS-VMS formulation with two variants: RPCS-VMS (No recovery stabilization) and RPCS-VMS (No increment stabilization). In each variant, the entire residual contribution is removed from the corresponding equation, including its terms on both sides of Eq.~\eqref{eq:vms-pressure-recovery} or Eq.~\eqref{eq:vms-pressure-increment}. All three formulations retain the same intermediate-velocity equation and its stabilization. We repeat the BDF2 manufactured-solution test of Section~\ref{subsec:manufactured-stability}.

\begin{figure}[!htbp]
    \centering
    \includegraphics[width=0.9\textwidth]{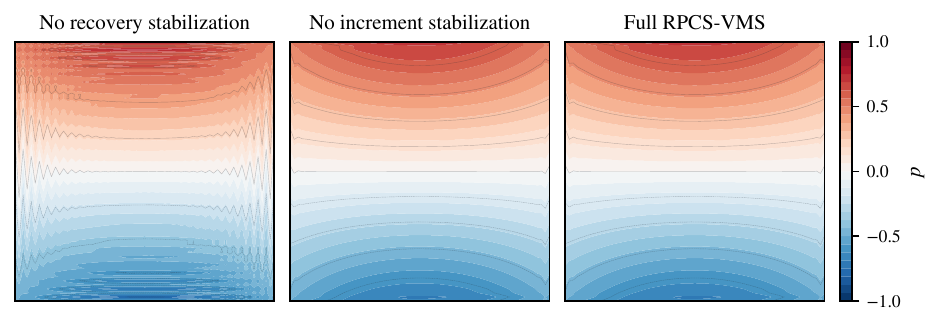}
    \caption{Pressure $p_h$ at $T=1$ for the manufactured-solution test with BDF2, $P_1/P_1$ elements, and $\Delta t=10^{-3}$. All three formulations retain rotational pressure recovery. All panels use the same color scale.}
    \label{fig:pressure_stage_ablation}
\end{figure}

Fig.~\ref{fig:pressure_stage_ablation} shows that removing pressure-recovery stabilization produces visible mesh-scale oscillations, whereas removing pressure-increment stabilization gives a pressure field visually similar to that of the full formulation. Table~\ref{tab:pressure_stage_ablation} reports the same $L^2$ error measures as Table~\ref{tab:boundary_pressure_comparison} and includes RPCS ($P_1/P_1$) and Monolithic ($P_2/P_1$) from Section~\ref{subsec:manufactured-stability}. Compared with the full RPCS-VMS scheme, removing pressure-recovery stabilization increases the global pressure error by a factor of $7.60$, with increases in both the boundary strip and the interior. Removing pressure-increment stabilization changes the global pressure error by less than $0.1\%$. Pressure-recovery stabilization therefore provides the main suppression of pressure oscillations in this test. Retaining both stabilization terms in the final algorithm is motivated by the coupled dissipation structure established for the coordinated formulation in Section~\ref{sec:stability}.

\par\addvspace{10pt}
\noindent\begin{minipage}{\linewidth}
\begin{center}
\begin{tabular}{@{}p{0.32\linewidth}cccc@{}}
\toprule
Method & $\|p-p_h\|_{\Omega}$ & $\|p-p_h\|_{\Omega_h^b}$ & $\|p-p_h\|_{\Omega_h^i}$ & $\|\bm u-\bm u_h\|_{\Omega}$ \\
\midrule
RPCS ($P_1/P_1$) & $4.1429\times10^{-2}$ & $1.4873\times10^{-2}$ & $3.8668\times10^{-2}$ & $2.8351\times10^{-4}$ \\
RPCS-VMS (No recovery stabilization) & $2.5573\times10^{-2}$ & $1.0828\times10^{-2}$ & $2.3167\times10^{-2}$ & $3.0302\times10^{-4}$ \\
RPCS-VMS (No increment stabilization) & $3.3626\times10^{-3}$ & $2.6642\times10^{-3}$ & $2.0516\times10^{-3}$ & $2.8223\times10^{-4}$ \\
RPCS-VMS ($P_1/P_1$) & $3.3659\times10^{-3}$ & $2.6720\times10^{-3}$ & $2.0470\times10^{-3}$ & $3.0222\times10^{-4}$ \\
Monolithic ($P_2/P_1$) & $3.0916\times10^{-5}$ & $1.3382\times10^{-5}$ & $2.7870\times10^{-5}$ & $8.9860\times10^{-7}$ \\
\bottomrule
\end{tabular}
\end{center}
\refstepcounter{table}
\label{tab:pressure_stage_ablation}
{\small
\noindent\textbf{Table~\thetable}\par
\noindent Unsteady manufactured-solution $L^2$ errors at $T=1$ with BDF2, $\Delta t=10^{-3}$, and $n=64$. The error definitions and boundary strip are the same as in Table~\ref{tab:boundary_pressure_comparison}. All RPCS formulations use $P_1/P_1$ elements. Monolithic uses unstabilized $P_2/P_1$ elements.\par
}
\end{minipage}\par
\par\addvspace{10pt}

A possible explanation follows from the structure of the rotational update. Without pressure-recovery stabilization, Eq.~\eqref{eq:vms-pressure-recovery} computes the $L^2$ projection of $p_h^n+\phi_h-\nu\nabla\cdot\tilde{\bm u}_h$ onto the pressure space. For the continuous $P_1$ velocity approximation used here, $\nabla\cdot\tilde{\bm u}_h$ is elementwise constant and can jump across element interfaces. Spatial variations in this divergence correction can therefore enter the pressure even when the pressure-increment equation is stabilized. The recovery stabilization adds the positive gradient term $\nu(\tau_m\nabla p_h,\nabla r_h)_h$ to the pressure operator, together with the associated momentum-residual forcing. This gradient term acts directly on the pressure and provides a mechanism for suppressing mesh-scale variations, consistent with the smoother fields observed when recovery stabilization is retained.

\FloatBarrier
\subsection{Branched Pipe Benchmark}

We next simulate the branched-pipe benchmark of Liu \citep{Liu2009Open}, also studied by Poux et al.~\citep{Poux2011Improvements}. This case examines separated internal flow with strong turning, a thin internal obstacle, and flux partition between two open outlets. The 2D domain is
\begin{equation}
    \Omega=[0,8]\times[-0.5,0.5]\setminus
    \{[0,0.5]\times[-0.5,0]\cup[1.5,8]\times[-0.1,0.2]\}.
\end{equation}
The inlet is the segment $x=0$, $0\le y\le0.5$. The open outlets are the right-end segments $x=8$, $0.2\le y\le0.5$ and $x=8$, $-0.5\le y\le-0.1$. All other boundaries are no-slip. The simulation uses $\nu=1/600$, BDF2 with the time step in Table~\ref{tab:numerical_reproducibility}, and zero pseudo-traction at the outlets. Following the transient start-up of Liu \citep{Liu2009Open}, the inlet velocity is $\bm u_{\rm in}(y,t)=s(t)(12y(1-2y),0)$, with $s(t)=(1-\cos(\upi t))/2$ for $0\le t\le1$ and $s(t)=1$ for $t\ge1$.

\begin{figure}[!htb]
    \centering
    \includegraphics[width=\textwidth]{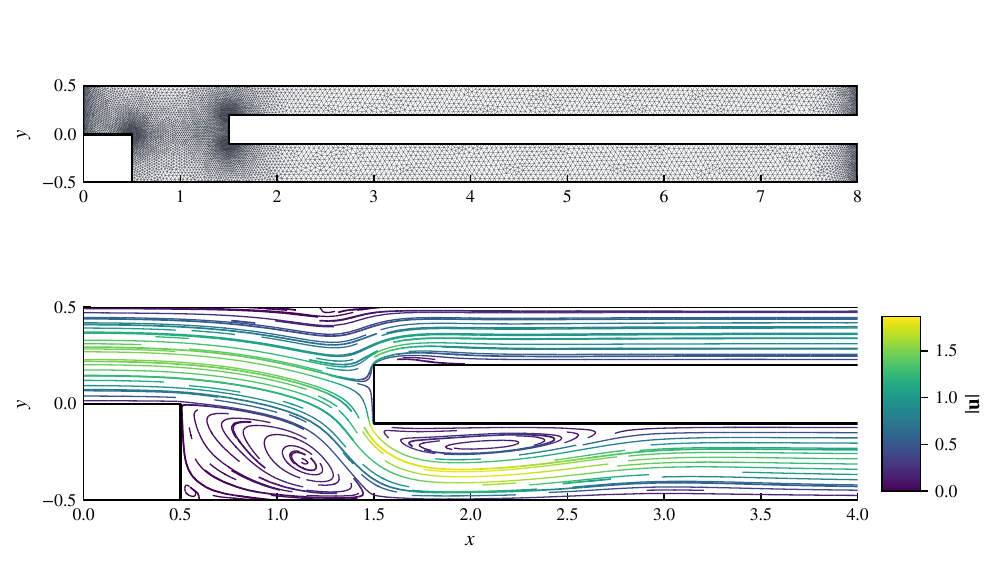}
    \caption{Branched-pipe mesh and flow structure. Top: computational mesh with local refinement near the inlet, outlets, entrance corner, and splitter tip. Bottom: streamlines showing jet deflection, lower-cavity recirculation, and downstream recovery.}
    \label{fig:splitter}
\end{figure}

\begin{figure}[!htb]
    \centering
    \includegraphics[width=0.75\textwidth]{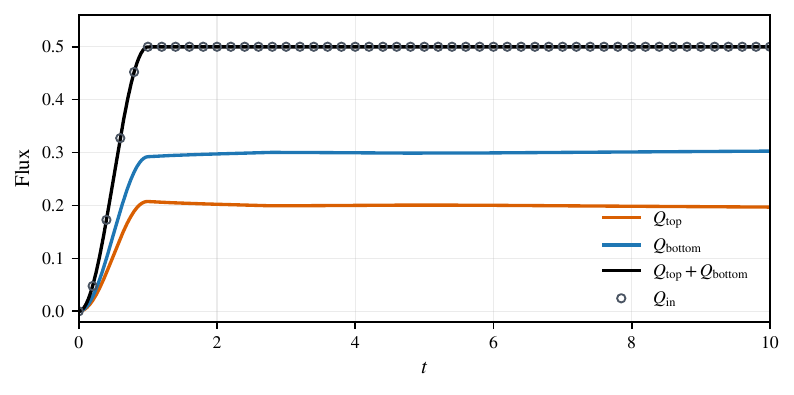}
    \caption{Branched-pipe flux histories during start-up. The upper and lower outlet fluxes show the flow partition, while their sum is compared with the prescribed inflow to assess global mass conservation.}
    \label{fig:bifurcated_outflux}
\end{figure}

The mesh in Fig.~\ref{fig:splitter} uses local refinement near the inlet, outlets, entrance corner, and splitter tip. The streamlines show jet deflection, lower-cavity recirculation, and downstream recovery. The start-up histories in Fig.~\ref{fig:bifurcated_outflux} show how the outlet fluxes develop over $0\le t\le10$. The analytic inflow is $Q_{\rm in}(t)=(1-\cos(\upi t))/4$ during start-up and $Q_{\rm in}=0.5$ afterwards. The summed outlet flux follows this prescribed inflow closely, indicating good global mass conservation. The individual fluxes show how the flow is partitioned between the two branches.

In addition to the start-up histories, Table~\ref{tab:bifurcated_tube_metrics} compares three integral diagnostics at $T=50$ against a Monolithic $P_2/P_1$ reference solution computed on the same resolution: the kinetic energy $e_c$, the outlet flux $Q_R$, and the inlet-to-outlet mechanical-energy drop $\Delta h_R$, referred to as the charge drop \citep{Poux2011Improvements}. These quantities are defined as
\begin{equation}
e_c=\int_{\Omega}\frac{\rho_0}{2}|\bm u|^2\,\mathrm{d}\bm x,
\qquad
Q_R=\int_R \bm u\cdot\bm n\,\mathrm{d}l,
\end{equation}
\begin{equation}
\Delta h_R
=\frac{1}{2}\left(
\int_L(\bm u\cdot\bm n)^2\,\mathrm{d}l
-\int_R(\bm u\cdot\bm n)^2\,\mathrm{d}l
\right)
+\left(\int_L p\,\mathrm{d}l-\int_R p\,\mathrm{d}l\right).
\end{equation}
Here, $\rho_0=1$ is the nondimensional density, and $L$ and $R$ denote the inlet and outlet boundaries, respectively.
RPCS-VMS agrees closely with the Monolithic $P_2/P_1$ reference in kinetic energy and branch fluxes. The charge drops differ only by about $1.5\%$ and $1.8\%$ at the top and bottom outlets.

\begin{table}[!htbp]
\centering
\begin{tabular}{lcc}
\toprule
Quantity & RPCS-VMS & Monolithic $P_2/P_1$ \\
\midrule
Total kinetic energy & 1.9941 & 1.9899 \\
Top outflux & 0.1934 & 0.1935 \\
Bottom outflux & 0.3063 & 0.3065 \\
Top charge drop & 0.6334 & 0.6238 \\
Bottom charge drop & 0.5674 & 0.5576 \\
\bottomrule
\end{tabular}
\caption{Branched-pipe integral diagnostics at $T=50$: total kinetic energy, upper and lower outlet fluxes, and inlet-to-outlet charge drops. RPCS-VMS is compared with the Monolithic $P_2/P_1$ reference at the same mesh resolution.}
\label{tab:bifurcated_tube_metrics}
\end{table}

\subsection{Flow Past a 2D Cylinder}

\begin{figure}[!htb]
    \centering
    \includegraphics[width=0.85\textwidth]{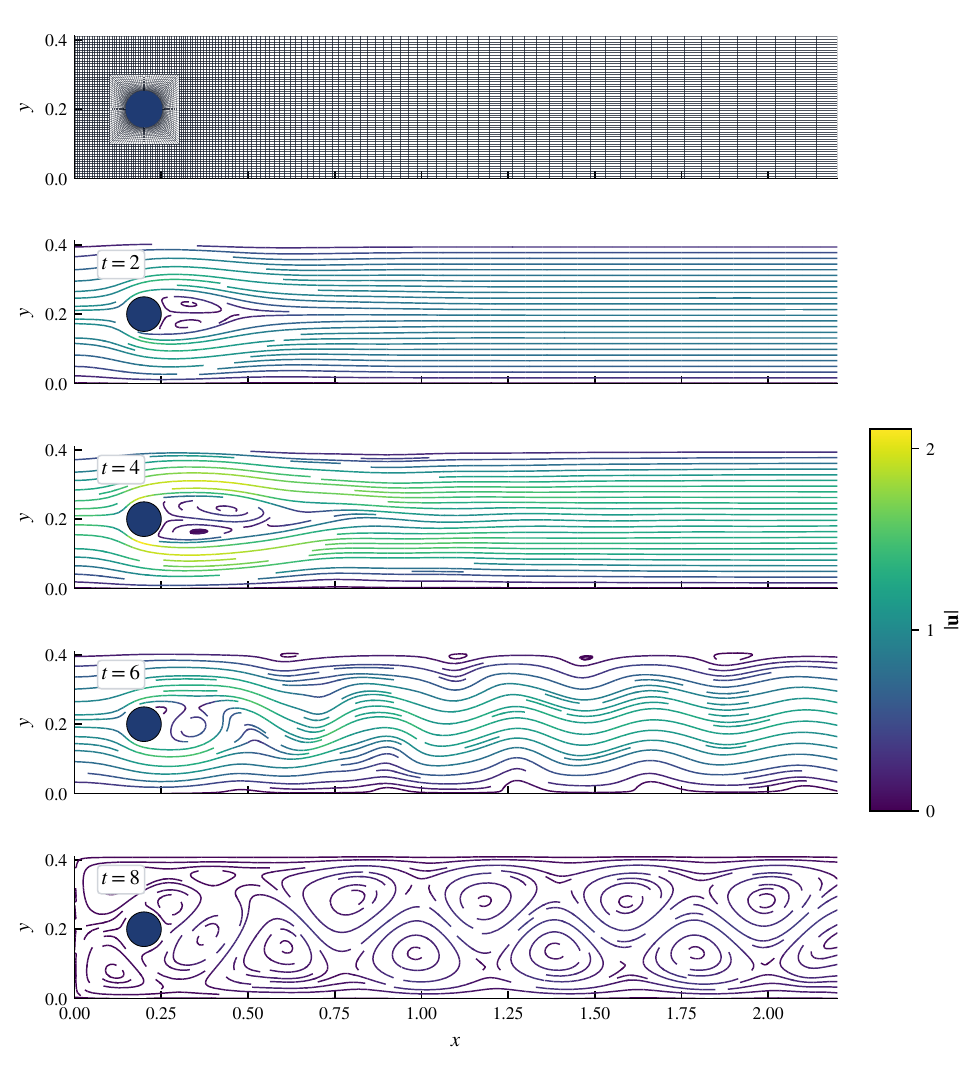}
    \caption{2D cylinder mesh and transient wake. Top: quadrilateral mesh. Below: streamlines colored by velocity magnitude.}
    \label{fig:cyl2d_stream}
\end{figure}

The 2D DFG cylinder benchmark complements the branched-pipe case by testing boundary-layer resolution, vortex shedding, pressure-drop prediction, and force integration \citep{FeatFlowDFG2D3,Schafer1996Benchmark,John2004Reference}. The domain is
\begin{equation}
    \Omega=[0,2.2]\times[0,0.41]
    \setminus\left\{(x-0.2)^2+(y-0.2)^2\le 0.05^2\right\}.
\end{equation}
The cylinder and channel walls are no-slip, $x=0$ is the inflow, and $x=2.2$ is the zero-traction outlet. The prescribed inflow is
\begin{equation}
    u_x(y,t)=6\sin(\upi t/8)\frac{y(0.41-y)}{0.41^2}, \qquad u_y=0.
\end{equation}
The simulation uses $\nu=10^{-3}$ and the BDF2 time step in Table~\ref{tab:numerical_reproducibility}. The pressure difference $\Delta p$ is computed between the points $(0.15,0.2)$ and $(0.25,0.2)$, while the force coefficients are normalized using the characteristic velocity $U=1$ and cylinder diameter $D=0.1$. Fig.~\ref{fig:cyl2d_stream} shows the quadrilateral mesh and the wake at four stages of the transient. The streamline fields trace the development of alternating wake vortices as the flow develops.

The drag and lift coefficients are evaluated on the cylinder boundary using the pseudo-stress convention consistent with the open-boundary formulation,
\begin{equation}
    C_i=\frac{2\nu}{U^2A_{\rm ref}}
    \int_{\Gamma_c}
    -\bm e_i\cdot
    \left(\nabla\bm u_h-\frac{p_h}{\nu}\bm I\right)\bm n\,\mathrm{d}s,
    \qquad i=x,y,
\label{eq:cylinder-force-coefficients}
\end{equation}
Here $\Gamma_c$ is the cylinder boundary, $\bm e_i$ the Cartesian unit vector, and $\bm I$ the identity tensor. The reference area is $A_{\rm ref}=D$ in 2D, and $C_D=C_x$ and $C_L=C_y$ denote drag and lift, respectively.

\begin{figure}[!htb]
    \centering
    \includegraphics[width=\textwidth]{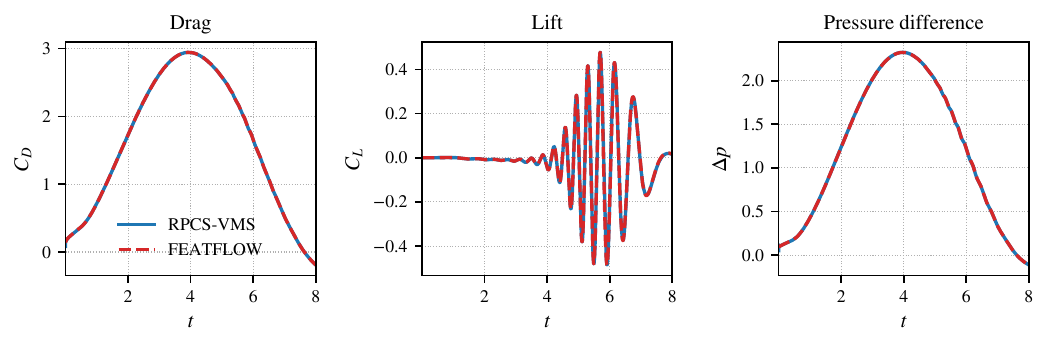}
    \caption{2D cylinder histories computed with RPCS-VMS and the FEATFLOW reference data \citep{John2004Reference,FeatFlowDFG2D3}.}
    \label{fig:cyl2d_comparison}
\end{figure}

Fig.~\ref{fig:cyl2d_comparison} compares the drag, lift, and pressure-difference histories with the FEATFLOW reference data \citep{John2004Reference,FeatFlowDFG2D3}. RPCS-VMS captures the rise and fall of drag and pressure difference, together with the growth and subsequent decay of the lift oscillations. The close agreement in the lift history indicates that both the amplitude and timing of the vortex-shedding response are reproduced. Table~\ref{tab:cylinder_2d_metrics} gives the quantitative comparison of some benchmark indices. Here $t(C_{D,\max})$ and $t(C_{L,\max})$ denote the times at which the drag and lift coefficients attain their respective maximum values over the simulated interval. All listed relative errors are within $0.18\%$, indicating excellent agreement.

\begin{table}[H]
\centering
\begin{tabular}{lccc}
\toprule
Metric & FEATFLOW & RPCS-VMS & Rel. err. (\%) \\
\midrule
$t(C_{D,\max})$ & 3.9362 & 3.9362 & 0 \\
$C_{D,\max}$ & 2.9438 & 2.9434 & 0.01 \\
$t(C_{L,\max})$ & 5.6925 & 5.6950 & 0.04 \\
$C_{L,\max}$ & 0.4775 & 0.4772 & 0.06 \\
$\Delta p(t=8)$ & -0.1115 & -0.1117 & 0.18 \\
\bottomrule
\end{tabular}
\caption{2D cylinder benchmark indices for RPCS-VMS and the FEATFLOW reference \citep{FeatFlowDFG2D3}, including peak force coefficients, their occurrence times, and the pressure difference at $t=8$. Relative errors are measured against the reference values.}
\label{tab:cylinder_2d_metrics}
\end{table}

\subsection{Flow Past a 3D Cylinder}

\begin{figure}[!htb]
    \centering
    \includegraphics[width=0.76\textwidth]{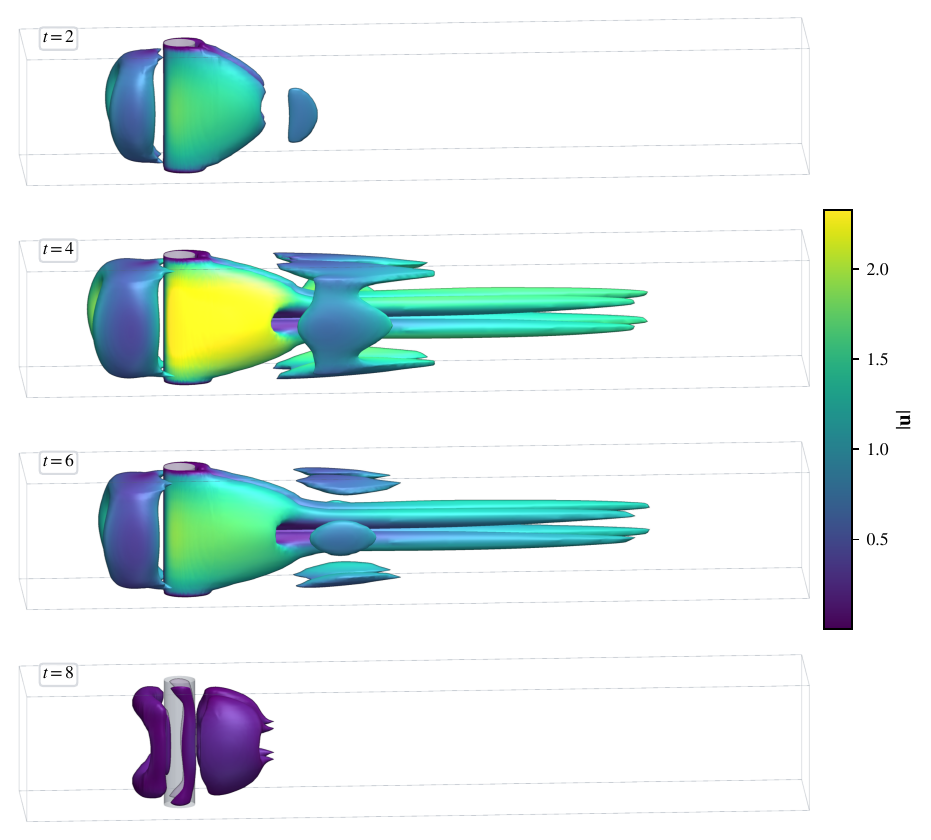}
    \caption{3D cylinder wake visualized by $Q$-criterion isosurfaces at the common threshold $Q=2.618$, colored by velocity magnitude.}
    \label{fig:cyl3d_mesh_q}
\end{figure}

As shown in Fig. \ref{fig:cyl3d_mesh_q}, the 3D DFG cylinder benchmark extends the force comparison to a 3D wake with a smaller lift signal \citep{Schafer1996Benchmark,BayraktarMierkaTurek2012Benchmark,FeatFlowDFG3DReference}. The domain is
\begin{equation}
    \Omega=[0,2.5]\times[0,0.41]\times[0,0.41]
    \setminus\left\{(x-0.5)^2+(y-0.2)^2\le 0.05^2,
    \ 0\le z\le0.41\right\}.
\end{equation}
The cylinder axis is parallel to $z$. The plane $x=0$ is the inflow, $x=2.5$ is the zero-traction outlet, and the walls and cylinder surface are no-slip. The simulation uses $\nu=10^{-3}$, the BDF2 time step in Table~\ref{tab:numerical_reproducibility}, and
\begin{equation}
    u_x(y,z,t)=36\sin(\upi t/8)
    \frac{yz(0.41-y)(0.41-z)}{0.41^4},
    \qquad u_y=u_z=0.
\end{equation}
The drag and lift coefficients are evaluated using Eq.~\eqref{eq:cylinder-force-coefficients} with $U=1$, $D=0.1$, and $A_{\rm ref}=0.041$.

\begin{figure}[!htb]
    \centering
    \includegraphics[width=0.92\textwidth]{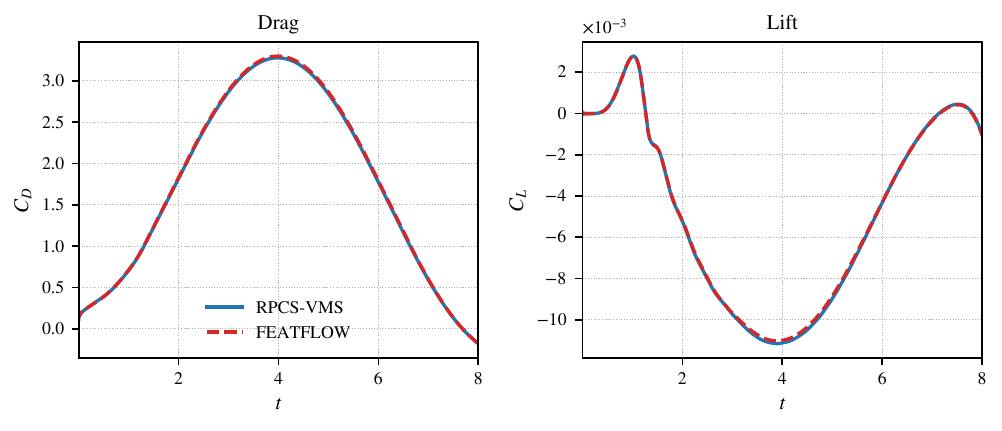}
    \caption{3D cylinder force histories computed with RPCS-VMS and the FEATFLOW reference data \citep{BayraktarMierkaTurek2012Benchmark,FeatFlowDFG3DReference}.}
    \label{fig:cyl3d_comparison}
\end{figure}

Fig.~\ref{fig:cyl3d_comparison} compares the force histories with the FEATFLOW reference data \citep{BayraktarMierkaTurek2012Benchmark,FeatFlowDFG3DReference}. RPCS-VMS reproduces the transient growth and decay of drag and the much smaller lift response. Table~\ref{tab:cylinder_3d_metrics} quantifies the agreement at the force extrema. The maximum drag differs by $0.62\%$, while the maximum and minimum lift differ by $0.28\%$ and $1.82\%$, respectively. Together, the histories and extrema show that both streamwise and transverse forces are accurately captured at this resolution.

\begin{table}[H]
\centering
\begin{tabular}{lccc}
\toprule
Metric & FEATFLOW & RPCS-VMS & Rel. err. (\%) \\
\midrule
$C_{D,\max}$ & 3.2978 & 3.2775 & 0.62 \\
$C_{L,\max}$ & $2.7755\times 10^{-3}$ & $2.7833\times 10^{-3}$ & 0.28 \\
$C_{L,\min}$ & -0.0110 & -0.0112 & 1.82 \\
\bottomrule
\end{tabular}
\caption{3D cylinder force extrema for RPCS-VMS and the FEATFLOW reference \citep{FeatFlowDFG3DReference}. Relative errors are measured against the reference values.}
\label{tab:cylinder_3d_metrics}
\end{table}

We further assess computational cost by comparing RPCS-VMS with Monolithic-VMS over the first 100 time steps ($T=0.5$) using the 3D cylinder case. Both use the same mesh, $\nu=10^{-3}$, $\Delta t=0.005$, $C_I=36$, and $\tau_c=\nu/6$. They share the time integration method and elementwise stabilization model. Convection is linearized using an extrapolated advecting velocity in both formulations. Each time step requires four sequential linear solves for RPCS-VMS and one coupled linear solve for Monolithic-VMS.

Each method is run three times using eight MPI ranks (one thread per rank) on an Intel Xeon Gold 6226R processor under Ubuntu 20.04. To facilitate a fair performance comparison with RPCS-VMS, the Monolithic-VMS system is solved using GMRES with a block Schur-complement preconditioner based on a lower-triangular factorization of the coupled velocity–pressure system. This preconditioner enables a sequential treatment of the velocity and pressure blocks, similar to the solution strategy of RPCS-VMS. The velocity and Schur-complement block solves are each approximated by BoomerAMG. Both formulations use the same relative and absolute solver convergence tolerances.

The total wall time includes assembly, preconditioning, and linear solves, but excludes initialization and field output. Averaged over three runs, the wall times are 5550.0 s for Monolithic-VMS and 2301.7 s for RPCS-VMS, corresponding to a 2.41-fold speedup for the reported configuration. These results highlight the computational advantage of the RPCS-VMS formulation, which replaces the coupled velocity-pressure solve with a sequence of smaller, decoupled systems, substantially reducing the overall computational cost in this case.

\section{Conclusion}
\label{sec:conclusions}

In this study, we propose a variational multiscale rotational pressure-correction method for incompressible flows with equal-order finite elements. The method incorporates residual stabilization in both pressure equations while preserving the four-solve sequential structure. For the reduced BDF1 Oseen model analyzed here, we establish augmented-energy decay and a conditional pressure-gradient bound under stated coefficient conditions.

Numerical experiments demonstrate good spatial and temporal convergence rates, mitigation of small-time-step pressure oscillations, and close agreement with reference solutions for separated flows. A comparison of the two pressure-stage stabilization terms indicates that recovery stabilization plays the principal role in suppressing oscillations in the tested case. In the 3D cylinder benchmark, RPCS-VMS achieves a 2.41-fold wall-time speedup relative to the tested Monolithic-VMS solver. These results support the use of RPCS-VMS as an efficient equal-order formulation for the incompressible flow problems considered here.

\section*{Code availability}

The source code accompanying this study is available through GitHub: \url{https://github.com/CZHU20/RPCS-VMS}.

\section*{Declaration of competing interest}
The authors declare that they have no known competing financial interests or personal relationships that could have appeared to influence the work reported in this paper.

\section*{Acknowledgements}
This work was supported by the National Key Research and Development Program of China (Grant No. 2021YFA1000201), the National Natural Science Foundation of China (Grant Nos. 12293001, 12272009), and the Beijing Natural Science Foundation (Grant No. L259063).

\section*{Declaration of generative AI and AI-assisted technologies in the manuscript preparation process}
The authors used Codex (OpenAI) to assist with language editing, manuscript organization, and checks of clarity and internal consistency. The authors evaluated the suggestions, revised the text, and verified the scientific content. The authors remain fully responsible for the final manuscript.

\appendix
\numberwithin{equation}{section}
\section{Proof of the Stability Estimates}
\label{app:energy-proof}

We derive the energy identity and pressure-gradient bound stated in Section~\ref{sec:stability}.

\subsection{Energy identity and decay}
\label{app:energy-identity}

For the piecewise linear fields used in this analysis, all elementwise second derivatives vanish. Under the assumptions of Section~\ref{subsec:stokes-energy-setting}, the intermediate velocity, pressure increment, velocity correction, and pressure recovery equations are
\begin{equation}
 (\widetilde{\bm a}_h^{n+1},\bm v_h)
 +(\mathcal C(\bm w_h^{n+1})\widetilde{\bm u}_h^{n+1},\bm v_h)
 +\nu(\nabla\widetilde{\bm u}_h^{n+1},\nabla\bm v_h)
 +(\tau_c \nabla\cdot\widetilde{\bm u}_h^{n+1},\nabla\cdot\bm v_h)_h
 -(p_h^n,\nabla\cdot\bm v_h)
 -\tfrac12(\beta^*\widetilde{\bm u}_h^{n+1},\bm v_h)_{\Gamma_N}=0,
\label{eq:stability-stage1}
\end{equation}
\begin{equation}
 (\nabla\cdot\widetilde{\bm u}_h^{n+1},q_h)+\Delta t(\nabla\phi_h^{n+1},\nabla q_h)
 +(\tau_m\bm R_\phi^{n+1},\nabla q_h)_h=0,
\label{eq:stability-increment}
\end{equation}
\begin{equation}
 \bm u_h^{n+1}
 =\widetilde{\bm u}_h^{n+1}-\Delta t\mathsf P_h\nabla\phi_h^{n+1},
\label{eq:stability-correction}
\end{equation}
\begin{equation}
 (p_h^{n+1}-\pi_h^{n+1}+\nu \nabla\cdot\widetilde{\bm u}_h^{n+1},r_h)
 +\nu(\tau_m\bm R_p^{n+1},\nabla r_h)_h=0,
\label{eq:stability-recovery}
\end{equation}
for \(\bm v_h\in X_h\), \(q_h\in M_{h,0}\), and \(r_h\in M_h\), with the momentum residuals given in Section~\ref{subsec:stokes-energy-setting}.

To combine the pressure-increment and pressure-recovery contributions, use the relations
\begin{equation}
\begin{aligned}
 \pi_h^{n+1}&=\psi_h^{n+1}-\eta_h^n,\qquad
 p_h^{n+1}=\pi_h^{n+1}-\delta\eta_h^{n+1},\qquad
 \bm R_p^{n+1}&=\bm R_\phi^{n+1}-\nabla\delta\eta_h^{n+1}.
\end{aligned}
\label{eq:stability-auxiliary-relations}
\end{equation}
Hence the pressure-recovery equation can be written as
\begin{equation}
 (\delta\eta_h^{n+1},r_h)
 =\nu(\nabla\cdot\widetilde{\bm u}_h^{n+1},r_h)
  +\nu(\tau_m\bm R_p^{n+1},\nabla r_h)_h,
 \qquad r_h\in M_h.
\label{eq:stability-eta-relation}
\end{equation}

The energy balance for the intermediate velocity follows by testing Eq.~\eqref{eq:stability-stage1} with
\(2\Delta t\widetilde{\bm u}_h^{n+1}\), using
\(2(x-y,x)=\|x\|^2-\|y\|^2+\|x-y\|^2\), and applying Eq.~\eqref{eq:stability-oseen-flux}:
\begin{equation}
\begin{aligned}
 &\|\widetilde{\bm u}_h^{n+1}\|^2-\|\bm u_h^n\|^2+T_n
 +2\nu \Delta t\|\nabla\widetilde{\bm u}_h^{n+1}\|^2\\
 &\qquad+2\Delta t\|\nabla\cdot\widetilde{\bm u}_h^{n+1}\|_{\tau_c}^2+\Delta t B_n
       -2\Delta t(p_h^n,\nabla\cdot\widetilde{\bm u}_h^{n+1})=0.
\end{aligned}
\label{eq:stability-intermediate-energy}
\end{equation}
Here \(T_n=\|\widetilde{\bm u}_h^{n+1}-\bm u_h^n\|^2\).
To replace intermediate by corrected kinetic energy, use orthogonality of \(\mathsf P_h\):
\[
(\widetilde{\bm u}_h^{n+1},\mathsf P_h\nabla\phi_h^{n+1})
=(\widetilde{\bm u}_h^{n+1},\nabla\phi_h^{n+1}).
\]
Integration by parts produces no boundary term:
\(\widetilde{\bm u}_h^{n+1}=0\) on \(\Gamma_D\) and
\(\phi_h^{n+1}=0\) on \(\Gamma_N\).
Therefore
\begin{equation}
 \|\bm u_h^{n+1}\|^2
 =\|\widetilde{\bm u}_h^{n+1}\|^2
  +2\Delta t(\nabla\cdot\widetilde{\bm u}_h^{n+1},\phi_h^{n+1})
  +(\Delta t)^2\|\mathsf P_h\nabla\phi_h^{n+1}\|^2.
\label{eq:stability-correction-norm}
\end{equation}
Substitution into Eq.~\eqref{eq:stability-intermediate-energy} combines the pressure contributions into
\(-2\Delta t(\pi_h^{n+1},\nabla\cdot\widetilde{\bm u}_h^{n+1})\).

To turn this pressure work into energy differences, test Eq.~\eqref{eq:stability-increment} with \(\psi_h^{n+1}\) and
Eq.~\eqref{eq:stability-eta-relation} with \(\eta_h^n\).
Using the pressure and residual relations in Eq.~\eqref{eq:stability-auxiliary-relations} and expanding the squared norms gives
\begin{equation}
\begin{aligned}
 -2\Delta t(\pi_h^{n+1},\nabla\cdot\widetilde{\bm u}_h^{n+1})
 ={}&(\Delta t)^2\bigl(\|\nabla\psi_h^{n+1}\|^2-\|\nabla\psi_h^n\|^2
                         +\|\nabla\phi_h^{n+1}\|^2\bigr)\\
 &+\frac{\Delta t}{\nu}\bigl(\|\eta_h^{n+1}\|^2-\|\eta_h^n\|^2
                         -\|\delta\eta_h^{n+1}\|^2\bigr)\\
 &+\Delta t\bigl(\|\nabla\eta_h^{n+1}\|_{\tau_m}^2
         -\|\nabla\eta_h^n\|_{\tau_m}^2
         -\|\nabla\delta\eta_h^{n+1}\|_{\tau_m}^2\bigr)\\
 &+2\Delta t(\tau_m\bm R_\phi^{n+1},\nabla\pi_h^{n+1})_h.
\end{aligned}
\label{eq:stability-pressure-transform}
\end{equation}
The weighted norm identity uses time-independence of \(\tau_m\).
Orthogonality also gives
\(\|\nabla\phi_h^{n+1}\|^2-\|\mathsf P_h\nabla\phi_h^{n+1}\|^2
=\|\mathsf Q_h\nabla\phi_h^{n+1}\|^2\).
Combining Eqs.~\eqref{eq:stability-intermediate-energy},
\eqref{eq:stability-correction-norm}, and
\eqref{eq:stability-pressure-transform} therefore gives
\begin{equation}
\begin{aligned}
 E^{n+1}-E^n+T_n+J_n
 &+\Delta t\mathscr D_h(\widetilde{\bm u}_h^{n+1})
  +\nu \Delta t\|\nabla\cdot\widetilde{\bm u}_h^{n+1}\|^2+\Delta t B_n\\
 &+2\Delta t(\tau_m\bm R_\phi^{n+1},\nabla\pi_h^{n+1})_h
  -\frac{\Delta t}{\nu}\|\delta\eta_h^{n+1}\|^2
  -\Delta t\|\nabla\delta\eta_h^{n+1}\|_{\tau_m}^2=0.
\end{aligned}
\label{eq:stability-exact-energy}
\end{equation}

It remains to combine the last three terms with the divergence contribution. Complete squares to expose their dissipation, retaining \(\Delta t B_n\).
Choose \(r_h=\delta\eta_h^{n+1}\) in Eq.~\eqref{eq:stability-eta-relation}.
Using Eq.~\eqref{eq:stability-auxiliary-relations} gives
\begin{equation}
 \frac1\nu\|\delta\eta_h^{n+1}\|^2
 +\|\nabla\delta\eta_h^{n+1}\|_{\tau_m}^2
 =(\nabla\cdot\widetilde{\bm u}_h^{n+1},\delta\eta_h^{n+1})
  +(\tau_m\bm R_\phi^{n+1},\nabla\delta\eta_h^{n+1})_h.
\label{eq:stability-eta-test}
\end{equation}
Expanding the two squares on the right below and applying Eq.~\eqref{eq:stability-eta-test} yields
\begin{equation}
\begin{aligned}
 &\nu\|\nabla\cdot\widetilde{\bm u}_h^{n+1}\|^2+\|\bm R_\phi^{n+1}\|_{\tau_m}^2
  -\frac1\nu\|\delta\eta_h^{n+1}\|^2
  -\|\nabla\delta\eta_h^{n+1}\|_{\tau_m}^2\\
 &\qquad=\frac1\nu\|\delta\eta_h^{n+1}-\nu \nabla\cdot\widetilde{\bm u}_h^{n+1}\|^2
              +\|\bm R_p^{n+1}\|_{\tau_m}^2.
\end{aligned}
\label{eq:stability-eta-squares}
\end{equation}
Finally, \(\bm R_\phi^{n+1}
=\widetilde{\bm a}_h^{n+1}+\nabla\pi_h^{n+1}\) implies
\begin{equation}
 2(\tau_m\bm R_\phi^{n+1},\nabla\pi_h^{n+1})_h
 -\|\bm R_\phi^{n+1}\|_{\tau_m}^2
 =\|\nabla\pi_h^{n+1}\|_{\tau_m}^2
  -\|\widetilde{\bm a}_h^{n+1}\|_{\tau_m}^2.
\label{eq:stability-residual-square}
\end{equation}
Substitution of Eqs.~\eqref{eq:stability-eta-squares} and
\eqref{eq:stability-residual-square} into Eq.~\eqref{eq:stability-exact-energy}
leaves precisely
\(E^{n+1}-E^n+
(T_n-\Delta t\|\widetilde{\bm a}_h^{n+1}\|_{\tau_m}^2)
+J_n+\Delta t F_n+\mathcal S_n+\Delta t B_n=0\).
The term in parentheses is \(W_n\), proving Eq.~\eqref{eq:stability-closed-step}.

To establish energy decay, we now show that each dissipation term is nonnegative.
Since \(\|\nabla\cdot\bm v_h\|^2\le d\|\nabla\bm v_h\|^2\),
\begin{equation}
\begin{aligned}
 \mathscr D_h(\bm v_h)
 &=2\nu\|\nabla\bm v_h\|^2
       +2\|\nabla\cdot\bm v_h\|_{\tau_c}^2
       -\nu\|\nabla\cdot\bm v_h\|^2\\
 &\ge\left[2\tau_c^{\min}-\left(1-\frac2d\right)\nu\right]
                  \|\nabla\cdot\bm v_h\|^2\ge0.
\end{aligned}
\label{eq:stability-dissipation-decomposition}
\end{equation}
The coefficient bound \(\tau_{m,K}\le \Delta t\) gives \(W_n\ge0\).
Together with the other nonnegative terms, Eq.~\eqref{eq:stability-closed-step} gives
\(E^{n+1}\le E^n\). Summing it proves Eq.~\eqref{eq:stability-global-energy}.

For the pressure estimate, positive definiteness at \(\kappa_c=0\) is also needed. Zero dissipation forces equality in the underlying pointwise Cauchy-Schwarz inequality, so \(\nabla\bm v_h|_K=\lambda_K\bm I\), where \(\lambda_K\) is constant on each linear element. Thus \(\bm v_h|_K=\lambda_K\bm x+\bm b_K\). Continuity across shared faces makes \(\lambda_K,\bm b_K\) common to all elements. The zero values on \(\Gamma_D\) then force \(\bm v_h=0\).

Guermond, Minev, and Shen \citep{Guermond2006Error} use a tunable coefficient \(\chi\) in the rotational update \(p^{n+1}=p^n+\phi^{n+1}-\chi\nabla\cdot\tilde{\bm u}^{n+1}\). For Laplacian viscosity with open boundaries, their sufficient coercivity condition is \(0<\chi<2\nu/d\). Here the rotational coefficient remains \(\nu\). The continuity-residual stabilization in Eq.~\eqref{eq:vms-continuity-subscale} supplies the additional divergence control, with the threshold in Eq.~\eqref{eq:stability-assumptions}. Its pressure-subscale construction is described in Section~\ref{subsec:vms-divergence-control}.

\subsection{Bound for the pressure gradient}
\label{app:energy-pressure}

To transfer pressure control through pressure recovery, we need divergence control and an inverse estimate:
\begin{equation}
 \|\nabla\cdot\bm v_h\|^2\le C_{d,h}\mathscr D_h(\bm v_h),
 \qquad
 \|\nabla r_h\|^2\le C_{\nabla,h}^2\|r_h\|^2,
 \qquad \bm v_h\in X_h,\quad r_h\in M_h.
\label{eq:stability-divergence-constant}
\end{equation}
Both constants are finite on the chosen mesh because \(\mathscr D_h\) is positive definite and the spaces are finite dimensional. To collect the coefficients in the pressure-gradient bound below, choose
\begin{equation}
\begin{aligned}
 C_{p,h}&=\max\{2,\ 4\nu\tau_m^{\max}C_{\nabla,h}^2,\
                       4\nu^2\tau_m^{\max}C_{\nabla,h}^2C_{d,h}\}.
\end{aligned}
\label{eq:stability-pressure-constants}
\end{equation}
The pressure-recovery dissipation already controls \(z_h^{n+1}=\delta\eta_h^{n+1}-\nu \nabla\cdot\widetilde{\bm u}_h^{n+1}\).
Applying the inverse estimate to \(\delta\eta_h^{n+1}\in M_h\) gives
\begin{equation}
\begin{aligned}
 \|\nabla\delta\eta_h^{n+1}\|_{\tau_m}^2
 &\le\tau_m^{\max}C_{\nabla,h}^2\|\delta\eta_h^{n+1}\|^2\\
 &\le2\tau_m^{\max}C_{\nabla,h}^2
       \bigl(\nu^2\|\nabla\cdot\widetilde{\bm u}_h^{n+1}\|^2+\|z_h^{n+1}\|^2\bigr)\\
 &\le2\tau_m^{\max}C_{\nabla,h}^2
       \bigl(\nu^2 C_{d,h}\mathscr D_h(\widetilde{\bm u}_h^{n+1})
                    +\|z_h^{n+1}\|^2\bigr).
\end{aligned}
\label{eq:stability-eta-gradient-bound}
\end{equation}
Since \(p_h^{n+1}=\pi_h^{n+1}-\delta\eta_h^{n+1}\),
the squared triangle inequality and Eq.~\eqref{eq:stability-eta-gradient-bound} give
\begin{equation}
\begin{aligned}
 \|\nabla p_h^{n+1}\|_{\tau_m}^2
 &\le C_{p,h}\left[
    \|\nabla\pi_h^{n+1}\|_{\tau_m}^2
    +\mathscr D_h(\widetilde{\bm u}_h^{n+1})
    +\nu^{-1}\|z_h^{n+1}\|^2\right].
\end{aligned}
\label{eq:stability-weighted-pressure}
\end{equation}
Multiply by \(\Delta t\), sum over the steps, and apply Eq.~\eqref{eq:stability-global-energy}. The bracketed terms are part of its nonnegative dissipation, so their \(\Delta t\)-weighted sum is at most \(E^0\). Finally, \(\tau_m^{\min}\|\nabla p_h^{n+1}\|^2\le\|\nabla p_h^{n+1}\|_{\tau_m}^2\) proves Eq.~\eqref{eq:stability-pressure-gradient}.

\subsection{Comparison without momentum-residual stabilization}
\label{app:energy-unstabilized}

For this comparison, set \(\tau_m=0\) in the four weak equations, with \(\tau_c\) prescribed independently at or above the divergence threshold. The energy identity remains valid with
\begin{equation}
 E_0^n=\|\bm u_h^n\|^2+(\Delta t)^2\|\nabla\psi_h^n\|^2
                    +\frac{\Delta t}{\nu}\|\eta_h^n\|^2,
 \qquad E_0^{n+1}\le E_0^n.
\label{eq:stability-energy-without-vms}
\end{equation}
However, its pressure-gradient dissipation is absent. Using \(p_h^{n+1}=\psi_h^{n+1}-\eta_h^{n+1}\) and the inverse estimate in Eq.~\eqref{eq:stability-divergence-constant} gives
\begin{equation}
\begin{aligned}
 (\Delta t)^2\|\nabla p_h^{n+1}\|^2
 &\le2(\Delta t)^2\|\nabla\psi_h^{n+1}\|^2
       +2(\Delta t)^2 C_{\nabla,h}^2\|\eta_h^{n+1}\|^2\\
 &\le2\max\{1,\nu\Delta t C_{\nabla,h}^2\}\,E_0^{n+1}
 \le2\max\{1,\nu\Delta t C_{\nabla,h}^2\}\,E_0^0.
\end{aligned}
\label{eq:stability-unstabilized-pressure-step}
\end{equation}
Dividing by \((\Delta t)^2\), multiplying by \(\Delta t\), and summing prove Eq.~\eqref{eq:stability-pressure-without-vms}.

To see why energy nonincrease does not ensure decay of every pressure mode, suppose there exists a nonconstant \(q_h\in M_h\), with zero mean in the closed case, such that
\[
 (q_h,\nabla\cdot\bm v_h)=0
 \qquad\text{for every }\bm v_h\in X_h.
\]
This mode produces no pressure contribution to the intermediate-velocity equation. With \(\tau_m=0\), the fields
\[
 \widetilde{\bm u}_h^{n+1}=\bm u_h^n=0,\qquad
 p_h^n=q_h,\qquad \phi_h^{n+1}=0
\]
satisfy Eqs.~\eqref{eq:stability-stage1}-\eqref{eq:stability-recovery} at every step. The auxiliary sequences are \(\psi_h^n=0\) and \(\eta_h^n=-q_h\), so \(E_0^n=(\Delta t/\nu)\|q_h\|^2\) is constant and satisfies \(E_0^{n+1}\le E_0^n\). Whenever such a mode exists, its pressure pattern therefore persists unchanged despite the nonincrease of energy.

\bibliographystyle{unsrtnat}
\renewcommand{\bibfont}{\fontsize{8pt}{8pt}\selectfont}
\bibliography{ref}

\end{document}